\documentclass[twocolumn]{aastex62} 

\hypersetup{linkcolor=red,citecolor=blue,filecolor=cyan,urlcolor=magenta}

\def\hi{\relax \ifmmode {\mbox H\,\textsc{i}}\else H\,{\scshape i}\fi}
\def\hii{\relax \ifmmode {\mbox H\,\textsc{ii}}\else H\,{\scshape ii}\fi}
\def\nii{\relax \ifmmode {\mbox N\,\textsc{ii}}\else N\,{\scshape ii}\fi}
\def\oi{\relax \ifmmode {\mbox O\,\textsc{i}}\else O\,{\scshape i}\fi}
\def\oii{\relax \ifmmode {\mbox O\,\textsc{ii}}\else O\,{\scshape ii}\fi}
\def\oiii{\relax \ifmmode {\mbox O\,\textsc{iii}}\else O\,{\scshape iii}\fi}
\def\cii{\relax \ifmmode {\mbox C\,\textsc{ii}}\else C\,{\scshape ii}\fi}
\def\ciii{\relax \ifmmode {\mbox C\,\textsc{iii}}\else C\,{\scshape iii}\fi}
\def\civ{\relax \ifmmode {\mbox C\,\textsc{iv}}\else C\,{\scshape iv}\fi}
\def\hei{\relax \ifmmode {\mbox He\,\textsc{i}}\else He\,{\scshape i}\fi}
\def\heii{\relax \ifmmode {\mbox He\,\textsc{ii}}\else He\,{\scshape ii}\fi}
\def\mgii{\relax \ifmmode {\mbox Mg\,\textsc{ii}}\else Mg\,{\scshape ii}\fi}
\def\sii{\relax \ifmmode {\mbox S\,\textsc{ii}}\else S\,{\scshape ii}\fi}
\def\siii{\relax \ifmmode {\mbox S\,\textsc{iii}}\else S\,{\scshape iii}\fi}
\def\hgi{\relax \ifmmode {\mbox Hg\,\textsc{i}}\else Hg\,{\scshape i}\fi}
\def\nai{\relax \ifmmode {\mbox Na\,\textsc{i}}\else Na\,{\scshape i}\fi}
\def\mgi{\relax \ifmmode {\mbox Mg\,\textsc{i}}\else Mg\,{\scshape i}\fi}
\def\caii{\relax \ifmmode {\mbox Ca\,\textsc{ii}}\else Ca\,{\scshape ii}\fi}

\received{2019 January 7}
\accepted{2019 July 7}

\submitjournal{ApJ}

\shorttitle{Characterizing the local $\rm SFR-Z_g$ relation in MaNGA spiral galaxies}
\shortauthors{S\'anchez-Menguiano et al.}

\begin{document}

\title{\uppercase{Characterizing the local relation between star formation rate and \\gas-phase metallicity in MaNGA spiral galaxies}}  

\email{lsanchez@iac.es}

\author{Laura S\'anchez-Menguiano} 
\affiliation{Instituto de Astrof\'isica de Canarias, La Laguna, Tenerife, E-38200, Spain}
\affiliation{Departamento de Astrof\'isica, Universidad de La Laguna, Spain}

\author{Jorge S\'anchez Almeida} 
\affiliation{Instituto de Astrof\'isica de Canarias, La Laguna, Tenerife, E-38200, Spain}
\affiliation{Departamento de Astrof\'isica, Universidad de La Laguna, Spain}

\author{Casiana Mu\~noz-Tu\~n\'on} 
\affiliation{Instituto de Astrof\'isica de Canarias, La Laguna, Tenerife, E-38200, Spain}
\affiliation{Departamento de Astrof\'isica, Universidad de La Laguna, Spain}

\author{Sebasti\'an F. S\'anchez}
\affiliation{Instituto de Astronom\'ia, Universidad Nacional Aut\'onoma de M\'exico, A.P. 70-264, C.P. 04510, M\'exico D.F., Mexico}

\author{Mercedes Filho} 
\affiliation{CENTRA-Ci\^encias, Science Faculty, University of Lisbon, 1749-016 Lisbon, Portugal}
\affiliation{Department of Physics Engineering, Faculty of Engineering, University of Oporto, 4200-465 Oporto, Portugal}

\author{Hsiang-Chih Hwang} 
\affiliation{Department of Physics \& Astronomy, Johns Hopkins University, Baltimore, MD 21218, USA}

\author{Niv Drory}
\affiliation{McDonald Observatory, The University of Texas at Austin, 1 University Station, Austin, TX 78712, USA}

\begin{abstract}
The role of gas accretion in galaxy evolution is still a matter of debate. The presence of inflows of metal-poor gas that trigger star formation bursts of low metallicity has been proposed as an explanation for the local anti-correlation between star formation rate (SFR) and gas-phase metallicity ($\rm Z_g$) found in the literature. In the present study, we show how the anti-correlation is also present as part of a diversified range of behaviours for a sample of more than 700 nearby spiral galaxies from the SDSS IV MaNGA survey. We have characterized the local relation between SFR and $\rm Z_g$ after subtracting the azimuthally averaged radial profiles of both quantities. 60\% of the analyzed galaxies display a $\rm SFR-Z_g$ anti-correlation, with the remaining 40\% showing no correlation (19\%) or positive correlation (21\%). Applying a Random Forest machine-learning algorithm, we obtain that the slope of the correlation is mainly determined by the average gas-phase metallicity of the galaxy. Galaxy mass, $g-r$ colors, stellar age, and mass density seem to play a less significant role. This result is supported by the performed 2nd-order polynomial regression analysis. Thus, the local $\rm SFR-Z_g$ slope varies with the average metallicity, with the more metal-poor galaxies presenting the lowest slopes (i.e., the strongest $\rm SFR-Z_g$ anti-correlations), and reversing the relation for more metal-rich systems. Our results suggest that external gas accretion fuels star-formation in metal-poor galaxies, whereas in metal-rich systems the gas comes from previous star formation episodes. 
\end{abstract} 



\section{Introduction} \label{sec:intro}
Dark matter over-densities in the early Universe become galaxies through a self-regulated process controlled by gas accretion and feedback from massive stars and black holes, as well as by galaxy mergers \citep[e.g.,][]{dekel2009, bouche2010, dave2011, dave2012, silk2012, lilly2013, sanchezalmeida2014}. 
Even though this picture is thought to be reliable, the emergence of fully fledged galaxies from the underlying physical laws is not yet properly understood. The formation and evolution of galaxies has to be modeled numerically using ad-hoc recipes for the key physical processes, since they cannot be computed self-consistently from first principles. This so-called {\em sub-grid physics} is tuned to reproduce some of the scaling properties observed in galaxies (e.g., the distribution of luminosities or stellar masses), whereas other scaling properties are used to support the consistency of the simulated galaxies
\citep[][]{ceverino2014, vogelsberger2014, crain2015, schaye2015, hopkins2014, hopkins2018, springel2018}.
Thus, the observed scaling relations are fundamental to assess the realism of our understanding of galaxy formation and evolution. 

Among these scaling relations, those that link star formation rates (SFRs) with properties of the star-forming gas provide direct information on the star-formation process and its regulation. It has been long known that the SFR of a galaxy scales with its stellar mass \citep[$M_\star$; e.g.,][]{brinchmann2004, daddi2007, noeske2007, renzini2015, canodiaz2016}. This so-called {\em main sequence} characterizes star-forming galaxies, which also follow a scaling relation linking the metallicity  of the gas involved in the ongoing star-formation ($Z_g$) with $M_\star$ \citep[the mass-metallicity relation: MZR hereinafter; e.g.,][]{skillman1989, tremonti2004, berg2012, sanchez2013}. Since both SFR and $Z_g$ increase with increasing $M_\star$, one may naively think that for a fixed $M_\star$, $Z_g$ increases with increasing SFR. However, 
\citet{mannucci2010} and \citet{laralopez2010} found that galaxies with a higher SFR  show lower $Z_g$ at a given $M_\star$. This relation between $M_\star$, SFR, and $Z_g$ is now called the fundamental metallicity relation (FMR), and it was suggested already in previous studies before the relation was coined \citep{ellison2008, peeples2009, lopezsanchez2010}. This relation could be explained in terms of cosmic gas accretion predicted in cosmological numerical simulations of galaxy formation \citep[e.g.,][and the references therein]{mannucci2010, brisbin2012, dave2012, maiolino2019}.

Most scaling relations were originally found considering global properties of galaxies. However, there is substancial evidence that they present a counterpart based on local variations, being suggested that the global relations could arise from the local ones \citep[e.g.][]{rosalesortega2012, barreraballesteros2016, canodiaz2016, hsieh2017, errozferrer2019}. This may also be the case with the FMR; various observational works have hinted that, within a given galaxy, star-forming regions of particularly high surface SFR are associated with drops in metallicity. Thus, for galaxies with the same $M_\star$, those holding more active star-forming regions would show larger integrated SFR and lower $Z_g$. 

Among the observational evidences, the extremely metal-poor galaxies of the local Universe studied by \citet{sanchezalmeida2015} have a dominant starburst with a metallicity between 5 and 10 times lower than the underlying host galaxy \citep[see also][]{sanchezalmeida2013, sanchezalmeida2014}. \citet{cresci2010} find an anti-correlation between SFR and metallicity in three galaxies at $z\sim3$.  A large \hii\ region at the edge of  J1411-00 contains most of the SFR of the galaxy, and has an oxygen abundance  $\sim$0.2 dex lower than the rest of the galaxy \citep{richards2014}. HCG~91c reveals that at least three \hii\ regions harbor an oxygen abundance $\sim$0.15~dex lower than expected from their immediate surroundings and from the abundance gradient \citep{vogt2015}. Using a sample of 14 star-forming dwarf galaxies in the local Universe, \citet{sanchezalmeida2018}  show the existence of a spaxel-to-spaxel anti-correlation between the index N2 (i.e., the ratio between $[{\rm N II}]\lambda 6583$ and ${\rm H}\alpha$) and the H$\alpha$ flux, which are usually employed as proxies for $Z_g$ and SFR, respectively. Thus, the observed N2 to H$\alpha$ relation may reflect a local anti-correlation between $Z_g$ and SFR. The galaxy sample employed by \citet{sanchezalmeida2018} was selected without a physically motivated criterium based only on the availability of the required integral field spectroscopic (IFS) data, therefore, the relation seems to reflect a trend common to dwarf galaxies. 

Based also on IFS data, \citet{hwang2019} study the existence of anomalously low metallicity (ALM) regions in 1222 star-forming galaxies in the SDSS IV MaNGA survey \citep{bundy2015}.  ALMs are defined as star-forming regions in which the gas-phase metallicity is anomalously-low compared to expectations from the empirical relation between metallicity and stellar surface mass-density at a given stellar mass. \citet{hwang2019} find ALMs in 25\,\% of the galaxies, therefore, it is a common phenomenon. The incidence rate of ALMs increases with both global and local specific SFR (i.e., ${\rm SFR}/M_\star$), and is higher in lower mass galaxies, in the outer regions of galaxies, and in morphologically disturbed galaxies. Even though the presence of ALMs does not directly prove the existence of a local anti-correlation between $Z_g$ and SFR, it is very suggestive since ALMs are both common and associated with enhanced star formation.  

In this paper, MaNGA galaxies are used for the first time to characterize the local relation between $Z_g$ and SFR. The large statistics of the survey, described in Sect.~\ref{sec:sample}, allows us to explore this relation in an extended range of galaxy masses not covered by previous studies. Firstly, we explain the methodology and the procedure followed to derive the properties of the star-forming regions (Sect.~\ref{sec:analysis}). Then we show that the relation is present in most star-forming galaxies (Sect.~\ref{sec:results1}). In the same section, we quantify the relation via the slope of a linear regression. This parameter is the basis for studying the dependence of the relation on galaxy properties, study that is carried out using tools of artificial intelligence, namely, random forests (Sect.~\ref{sec:results2}). In order to characterize such dependences, a complementary polynomial regression analysis is also carried out (Sect.~\ref{sec:results3}). Finally, the discussion of the results and the main conclusions are given in Sect.~\ref{sec:discussion}.

\section{Data and sample} \label{sec:sample}

\subsection{MaNGA data}
Mapping Nearby Galaxies at Apache Point Observatory \citep[MaNGA,][]{bundy2015} is an ongoing survey developed as part of the Sloan Digital Sky Survey IV (SDSS-IV) project \citep{blanton2017}. The aim of MaNGA is to obtain IFS information for a sample of 10\,000 galaxies up to $z\sim0.15$. The observations are conducted on the basis of an integral field unit (IFU) fiber system feeding the BOSS spectrographs \citep{smee2013} on the Sloan 2.5m telescope at Apache Point Observatory \citep[New Mexico,][]{gunn2006}. The IFU fibers are distributed in 17 hexagonal bundles with 5 different configurations that comprise from 19 to 127 fibers. The field of view (FoV) of the instrument varies from $12.5''$ to $32.5''$ in diameter \citep{drory2015}, with uniform coverage of the targets achieved thanks to the performed three-point dither pattern \citep{law2015}. The spectrographs provide a wavelength coverage from $3600$ \AA\ to $10300$ \AA, with a nominal resolution of $\lambda/\Delta\lambda \sim 2100$ at 6000 \AA\ \citep{smee2013}. 

The MaNGA mother sample is split into two main subsets, named {\it Primary} and {\it Secondary} samples, defined by two radial coverage goals \citep{wake2017}. The Primary sample comprises $\sim 5000$ galaxies reaching out to $1.5$ effective radii ($R_e$), and the Secondary one accounts for $\sim 3300$ objects with a coverage up to $2.5\,R_e$. An additional third subsample, known as the {\it Color-Enhanced} supplement ($\sim 1700$ galaxies), is selected to increase the number of systems that are underrepresented in the color-magnitude space (namely high-mass blue galaxies, low-mass red galaxies, and ``green valley'' galaxies).

MaNGA was designed to have the same number of galaxies in each bin of $i$-band absolute magnitude M$_i$ \citep{wake2017}. M$_i$ is a proxy for stellar mass, therefore, the survey roughly contains the same number of galaxies per log mass bin in the range between $10^9$ and $10^{11}$ M$_\odot$ (see Fig.~\ref{fig1}, top panel). Since stellar mass is one of them main drivers in setting galaxy properties \citep[e.g.][]{blanton2009}, MaNGA is a good starting point for any exploratory study such as ours.

The data analyzed here were calibrated with version 2.4.3 of the Data Reduction Pipeline \citep[DRP,][]{law2016}, that includes the standard steps for fiber-based IFS data reduction such as bias subtraction and flat-fielding, flux and wavelength calibration, or sky subtraction. At the end the pipeline provides a regular-grid datacube, with the first two coordinates indicating the right ascension and declination of the target and the third one being the step in wavelength. As a result, we have individual spectra for each sampled spaxel of $0.5'' \times 0.5''$ and a final spatial resolution of FWHM $\sim2.5''$, which corresponds to a physical resolution of $\sim 1.5$ kpc (at an average redshift of 0.03, see Sect.~\ref{subsec:sample} for details on the selected subsample).

More detailed information about the MaNGA sample, survey design, observational strategy, and data reduction can be found in \citet{law2015}, \citet{yan2016}, \citet{law2016}, and \citet{wake2017}.

\subsection{Analyzed sample}\label{subsec:sample}
In this study, the analyzed sample is extracted from the internal release MaNGA Product Launches 7 \mbox{(MPL-7)} comprising 4688 galaxies. MPL-7 is identical to SDSS-DR15 released in December 2018 \citep{aguado2018}. 

From the MPL-7, designed to have the same number of galaxies in each stellar mass bin based on the MaNGA sample definition, we selected galaxies following these additional criteria:
\begin{enumerate}
\item Galaxies with $z<0.05$ and observed with the 91- and 127- fiber bundles (corresponding to FoV diameters of $27.5''$ and $32.5''$, respectively). 
\item Galaxies with $b/a > 0.35$, corresponding to an inclination lower than approximately 70$\degr$. The axial ratios ($b/a$) were taken from the NASA Sloan ATLAS (NSA) v1\_0\_1 catalog and measured on the SDSS $r$-band images \citep{blanton2011}. 
\item Galaxies with morphological types $T\geq1$, corresponding to Sa types and later. Morphological classifications were adopted from the MaNGA Deep Learning Morphology Value Added catalog (MDLM-VAC) described in \citet{fischer2019}. The MDLM-VAC is a morphological catalog of MaNGA DR15 galaxies obtained with Deep Learning models trained and tested on SDSS-DR7 images \citep[see][]{dominguezsanchez2018}. All the $T$-Type values in the catalog have been eye-balled, and modified if necessary, for additional reliability \citep{fischer2019}.
\item Finally, we discarded galaxies for which the quality of the final DRP products (3D dacubes) did not meet quality standards because of potential issues during processing, such as the presence of many dead fibers or a critical failure in one or more frames (i.e., bad flags in the field DRP3QUAL contained in the output FITS headers of the datacubes, see \citealt{law2016} for details). 
\end{enumerate}

The first criterion was applied to restrict the study to large galaxies with a good spatial resolution and sampling coverage. The second one was aimed at avoiding problems due to projection effects, removing edge-on systems. The morphological criterion $T\geq1$ allowed us to preferentially select star-forming systems. Finally, the last request guaranteed the quality of the analyzed data. Cluster galaxies and close pairs have not been excluded from the sample under the assumption of having a negligible effect on the results. Although the galaxy environment strongly determines the balance between morphological types, studies agree that the effect on the scaling relations of galaxies with the same morphological type is very secondary \citep[see review by][]{blanton2009}.

After applying all these selection criteria, our sample comprises 821 galaxies. At the end, from all these objects, only those containing at least 10 star-forming spaxels are considered for the analysis, constituting a final sample of 736 galaxies. This number is selected to properly measure the radial profiles that we need to subtract in order to characterize local property variations (see Sect.~\ref{subsec:analysis2} for details in the definition of these star-forming spaxels).

\begin{figure}
\begin{center}
\resizebox{\hsize}{!}{\includegraphics{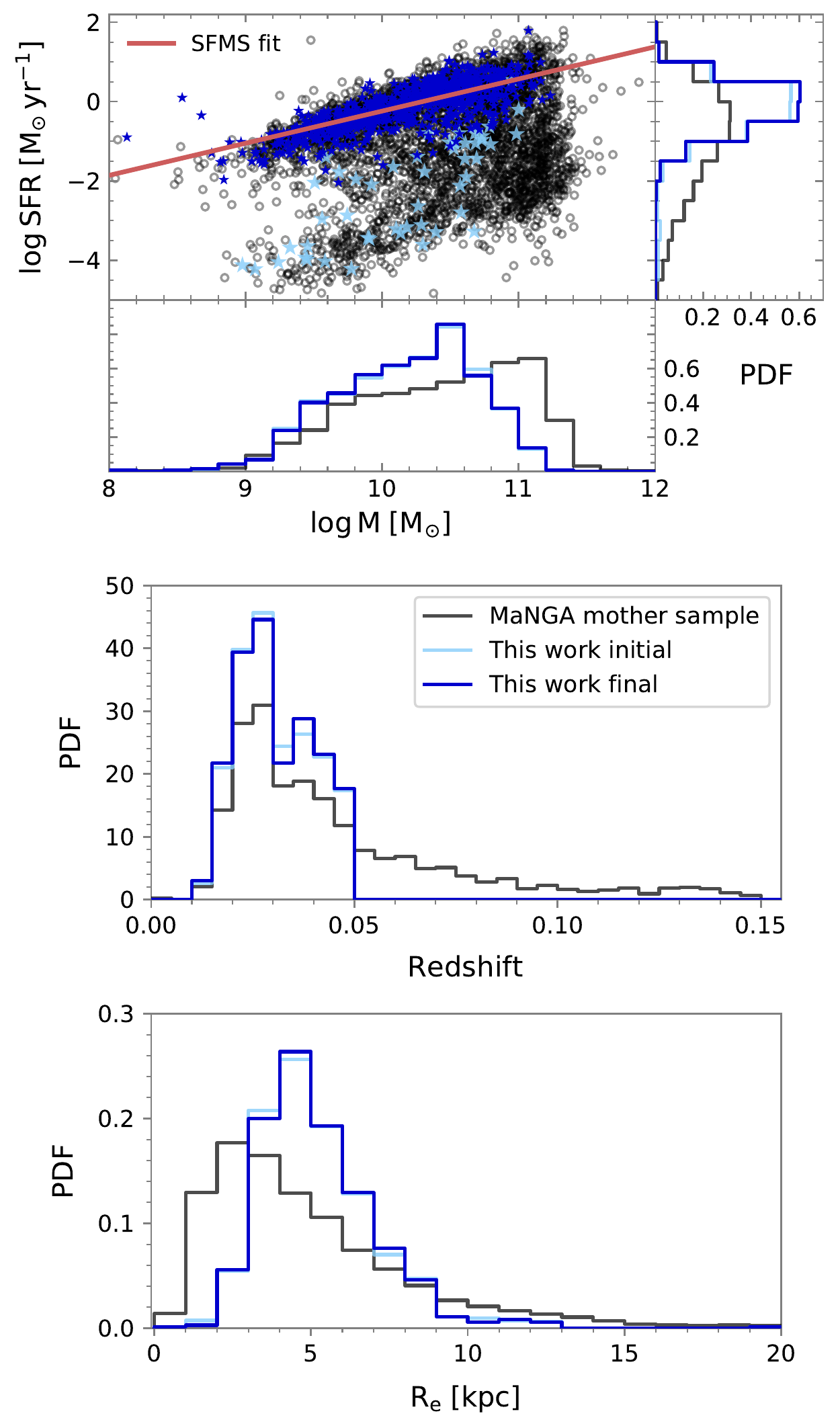}}
\caption{Star formation rate versus integrated stellar mass relation together with their probability distribution functions (PDF) for both MaNGA mother sample (black) and the analyzed subsample (blue, {\it top panel}). The PDF of redshifts ({\it middle}) and disc effective radii ({\it bottom}) are also shown. Dark blue represents the final selected sample, whereas light blue includes galaxies that were later discarded for insufficient number of star-forming spaxels. Solid line in top panel represents the linear fit to the star formation main sequence (SFMS) derived by \citet{canodiaz2016}.}
\label{fig1}
\end{center}
\end{figure}

The top panel of Fig.~\ref{fig1} shows the star formation rate-stellar mass plane of the galaxies in the sample (blue stars), in comparison with the MaNGA mother sample (black circles), representative of all types of galaxies observed in the nearby Universe. In this diagram we can see that the analyzed sample contains mainly galaxies occupying the ``star formation main sequence'' (SFMS), and a few systems falling in the ``green valley'' zone. The scarce objects of the sample corresponding to retired galaxies were those finally discarded for insufficient number of star-forming spaxels (light blue stars). The top panel also shows the  probability distribution functions (PDF) of both galaxy properties, where we can see the lack of high-mass early-type galaxies, with low SFRs. The middle panel of Fig.~\ref{fig1} presents the PDF of redshifts, where we can see a very similar distribution for both the MaNGA mother sample  (black) and the selected sample  (blue) below the truncation of the latter at $z=0.05$. Finally, the bottom panel of Fig.~\ref{fig1} shows the PDF of $R_{e}$ (extracted from the NSA catalog and measured on the SDSS $r$-band images), again for both this work and the MaNGA mother sample. We can see that both distributions are very similar except for the absence of the lowest end of effective radii, that is not covered because of the decision mentioned before of selecting only the two instrument configurations with the largest FoVs (the 91- and 127- fiber bundles), in order to limit the study to the galaxies with the best spatial resolution and sampling coverage. However, as we have seen in the top panel, this restriction on the effective radii does not produce any obvious bias on the sample.

\section{Analysis} \label{sec:analysis}

\subsection{Measurement of emission lines with {\scshape Pipe3D}}
In this study, we make use of the {\scshape Pipe3D} analysis pipeline \citep{sanchez2016a, sanchez2016b}, developed to characterize the properties of both the stellar populations and the ionized gas. This tool was designed to deal with spatially resolved data from optical IFS instruments. Its current implementation for the MaNGA data is described in more detail in \citet{sanchez2019}. 

Briefly, {\scshape Pipe3D} fits each spectrum by a linear combination of synthetic stellar population (SSP) templates \citep[following][]{cidfernandes2013} after correcting for the appropriate systemic velocity and velocity dispersion, and taking into account the effects of dust attenuation \citep{cardelli1989}. The SSP model spectra are then subtracted from the original cube to create a cube comprising only the ionized gas emission. In these spectra, {\scshape Pipe3D} measures the emission line fluxes performing a multi-component fitting using both a single Gaussian function per emission line and spectrum, and also a weighted moment analysis, as described in \citet{sanchez2016b}. A total of 52 emission lines are analyzed, obtaining the flux intensity, equivalent width (EW), systemic velocity, and velocity dispersion for each of them (including for instance H$\alpha$, H$\beta$, \mbox{[\oii]~$\lambda3727$}, \mbox{[\oiii]~$\lambda4959$}, \mbox{[\oiii]~$\lambda5007$}, \mbox{[\nii]~$\lambda6548$}, \mbox{[\nii]~$\lambda6584$}, \mbox{[\sii]~$\lambda6717,$} or \mbox{[\sii]~$\lambda6731$}). 

\subsection{Selection of star-forming regions} \label{subsec:analysis2}
The two-dimensional (2D) emission line intensity maps provided by {\scshape Pipe3D} are then corrected for dust attenuation making use of the extinction law from \citet{cardelli1989}, with $R_V=3.1$, and the H$\alpha$/H$\beta$ Balmer decrement, considering the theoretical value for the unobscured H$\alpha$/H$\beta$ ratio of 2.86, which assumes a case B recombination ($T_e = 10^4$ K, $n_e=10^2$ cm$^{-3}$, \citealt{osterbrock1989}). 

The relative error of the line flux intensities is estimated from the ratio of the flux to the flux error. We select spectra where this ratio is larger than 3 for each of the emission lines employed to derive the metallicities (see Sect.~\ref{subsec:metallicity}). We further select star-forming regions (spaxels) using the diagnostic BPT diagram proposed by \citet{baldwin1981}, based on the \mbox{[\nii]~$\lambda6584$/H$\alpha$} and \mbox{[\oiii]~$\lambda5007$/H$\beta$} line ratios. For this diagram we adopt the \citet{kewley2001} demarcation line to select our star-forming spaxels (those located below this curve), and having an $\rm H\alpha$ equivalent width greater than $\rm 6 \,\AA$. This second criterion assures the exclusion of low-ionization sources \citep[such as weak AGNs or post-AGB stars,][]{cidfernandes2011}, and the presence of a significant percentage ($\geq 20\%$) of young stars contributing to the emission of the star-forming regions \citep[given the strong correlation between both parameters, see][]{sanchez2014}. We note that using the \citet{kewley2001} curve instead of the one proposed by \citet{kauffmann2003} to select the star-forming spaxels does not produce any significant bias in the results.

\subsection{Metallicity}\label{subsec:metallicity}
As a proxy for $\rm Z_g$, we use the oxygen abundance (O/H) of the selected star-forming spaxels. $\rm Z_g$ is the fraction of metals by mass, whereas O/H is defined as the abundance of O, by number, relative to H. However, there is no inconsistency in using any of them to infer the local $\rm SFR-Z_{\lowercase{g}}$ relation since O/H and $\rm Z_g$ are proportional to each other, provided that the relative abundance of the different metals is the same for all galaxies.

The oxygen abundances are derived adopting the empirical calibration proposed by \citet[][hereafter M13]{marino2013} for the O3N2 index:
\begin{equation}
12+\log\left({\rm O/H}\right) = 8.533 - 0.214 \,\times\, {\rm O3N2}
,\end{equation}
with $\rm O3N2 =  \log\left([\oiii] \lambda5007/H\beta \times H\alpha/[\nii] \lambda6584\right)$. This relation is valid for the interval $-1.1 < {\rm O3N2} < 1.7$ (corresponding to $\rm 8.17 < 12+\log(O/H) < 8.77$). Therefore, star-forming regions presenting values outside this range are excluded from the analysis (representing barely 3\,\% of the detected \hii\,regions for 99.96\,\% of the galaxies).

The M13 calibration constitutes one of the most accurate calibrations to date for the O3N2 index, since it employs $T_e$-based abundances of $\sim 600$ \hii\,regions from the literature combined with new measurements from the \mbox{CALIFA} survey. The improvement of this calibration is especially significant in the high-metallicity regime, where previous calibrators based on this index lack high quality observations \citep[e.g.][]{pettini2004, perezmontero2009}. 

\begin{figure*}
\begin{center}
\resizebox{\hsize}{!}{\includegraphics{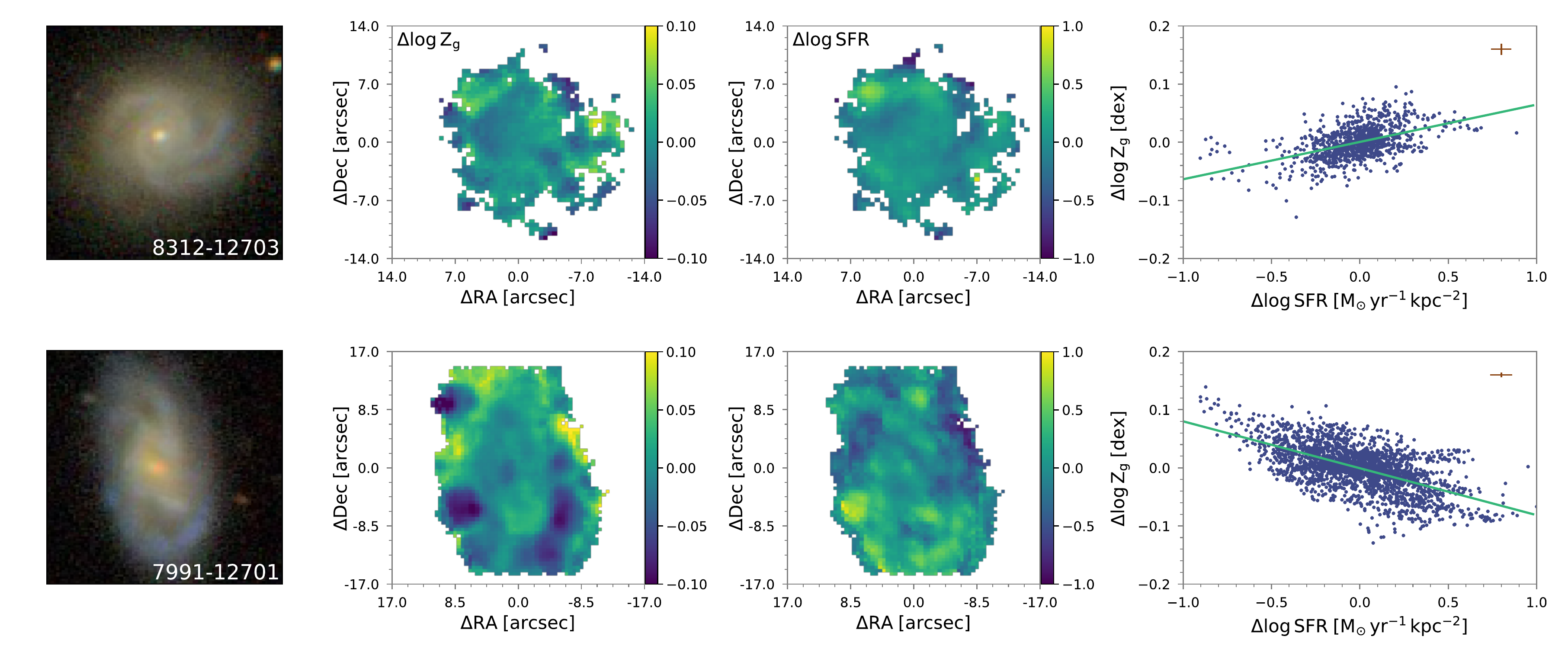}}
\caption{2D maps of the local behaviour of gas-phase metallicity ({\it middle left}) and SFR ({\it middle right}) after subtracting their radial profiles in two example galaxies (SDSS colour images of the galaxies are shown in the {\it left} panel, MaNGA IDs given on the right bottom corner). The relation between both parameters (expressed in logarithmic scale) is represented by a scatter plot ({\it right}), where the green solid line corresponds to a linear fit to such relation. Error bars indicated in the top right corner correspond to the average error of both residuals $\rm \Delta \log Z_g$ and $\rm \Delta \log SFR$. Note that the calibration error associated to the use of the O3N2 metallicity indicator is not represented (although it is included in further error propagations, see main text). Only the analyzed star-forming spaxels are represented in the figure.}
\label{fig2}
\end{center}
\end{figure*}

We also determine the oxygen abundance using version v.3.0 of HII-CHI-MISTRY \citep{perezmontero2014} in order to show the robustness of our results, that are not contingent upon the adopted method to measure the metallicity. This calibrator is based on a grid of photoionization models calculated using the {\tt CLOUDY} v.13 code \citep{ferland2013}. The grid covers a wide range of input conditions of abundances and ionization parameters, leading to the derivation of abundances consistent with the $T_e$-method based on collisionally excited lines. This calibration can be used through a publicly available Python module\footnote{\url{https://www.iaa.csic.es/\~epm/HII-CHI-mistry-opt.html}}. In this work, information on the [\oiii]$\lambda 4363$ emission line has not been fed to the code due to its faintness and consequently challenging detection. In this case, the algorithm uses a ``log U limited'' grid of photoionization models to derive reliable oxygen abundances \citep[for more information, see sect.4.2 of][]{perezmontero2014}.

\subsection{Star formation rate}\label{subsec:sfr}
The SFR is derived for the star-forming spaxels of each galaxy in the sample adopting the classical approach by \citet{kennicutt1998} based on the dust-corrected H$\alpha$ luminosities. These luminosities are obtained deriving the apparent magnitudes from the flux density in an H$\alpha$ image recovered from the data and then transforming these quantities to luminosities knowing the galaxy distance. Finally, in order to derive the SFR from H$\alpha$ luminosities we make use of the updated calibration presented in \citet{hao2011} for a Salpeter initial mass function \citep[IMF;][]{salpeter1955}:
\begin{equation}
{\rm SFR} \; [{\rm M}_{\odot}\,{\rm yr}^{-1}] = 8.79 \,\times\, 10^{-42} \, L\left({\rm H\alpha}\right) \, [\,{\rm erg}\,{\rm s}^{-1}]
.\end{equation} 

\subsection{Other global galaxy properties}
The total $M_\star$ of the galaxies is measured by {\scshape Pipe3D} from the SSP model spectra. This model is used to derive the stellar mass density ($\Sigma_\star$) of each spaxel adopting a Salpeter IMF, that is then co-added to estimate the integrated $M_\star$ of each galaxy with a typical error of 0.15 dex \citep{sanchez2016b}. 

The specific star formation rate (sSFR) is straightforwardly derived by dividing the total SFR of the galaxies from their integrated dust-corrected H$\alpha$ luminosities (determined in Sec.~\ref{subsec:sfr}) by their total $M_\star$.

In addition, as a result of the stellar continuum fit, {\scshape Pipe3D} provides 2D stellar velocity dispersion maps, as well as luminosity-weighted (LW) stellar age and metallicity maps from the individual weights of the combined single SPs. We use the mean LW stellar age and metallicity values measured at one effective radius as characteristic of the population of the entire galaxy. The central velocity dispersion ($\sigma_{cen}$) is estimated within an inner aperture of $2.5''$.

Another quantity analysed in this study is the $g-r$ color of the galaxies, that is derived using the model magnitudes in both bands corrected from extinction, obtained through a SQL query of the SDSS-DR14 database server\footnote{\url{http://skyserver.sdss.org/dr14/en/tools/search/sql.aspx}}.

Finally, the morphological information of the galaxy sample is extracted from \citet{fischer2019}. The published catalogue is based on Deep Learning techniques applied to SDSS-DR7 (see Sect.~\ref{subsec:sample} for more details).

\section{Existence of a local $\rm SFR-Z_{\lowercase{g}}$ relation}\label{sec:results1}
The aim of this study is to find and then characterize the existence of a local relation between SFR and metallicity. In order to describe local spatial variations of these properties, their characteristic radial profiles must be removed. For each individual galaxy, we derive the {\it residuals} by subtracting the azimuthally averaged radial $\rm Z_g$ (using the M13 calibration) and SFR distributions to the 2D maps. These averaged values are measured in elliptical annuli of $1''$ taking into account the position angle and ellipticity of the galaxies in order to correct for inclination. The resulting residuals expressed in logarithmic scale (hereafter $\rm \Delta \log Z_g$ and $\rm \Delta \log SFR$, respectively) for two example galaxies in the sample are shown in the left and middle panels of Fig.~\ref{fig2}. In the first case (8312-12703, top row), we can see that the peaks (yellow areas) and dips (dark blue) of both distributions are spatially coincident, whereas in the latter (7991-12701, bottom row) the highest metallicity residuals correspond to the lowest $\rm \Delta \log SFRs$.

This behaviour can be observed more clearly by representing $\rm \Delta \log Z_g$ of each star-forming spaxel against its $\rm \Delta \log SFR$. The right panels of Fig.~\ref{fig2} show this scatter plot for the two example galaxies. As hinted by the apparent coincidences in the 2D maps, the top galaxy exhibits a positive correlation between both properties, whereas the bottom galaxy shows a clear anti-correlation. 

In order to quantify the correlation, we perform a linear regression to the data points for each individual galaxy. The obtained slopes together with their corresponding errors are presented in Table~\ref{tab:app1} (Appendix~\ref{sec:app1}). Slope errors are derived via 100 Monte Carlo simulations taking into account the errors of both parameters $\rm \Delta \log Z_g$ and $\rm \Delta \log SFR$, that are determined from the propagation of the measurement errors of the emission line fluxes. For the metallicity, we have also propagated the calibration error associated to the scatter in the O3N2 relation \citep[0.08 dex, see][]{marino2013}, although it is not included in the average error bars represented in the top right corner of the right panels of Fig.~\ref{fig2}. 

\begin{figure}
\begin{center}
\resizebox{\hsize}{!}{\includegraphics{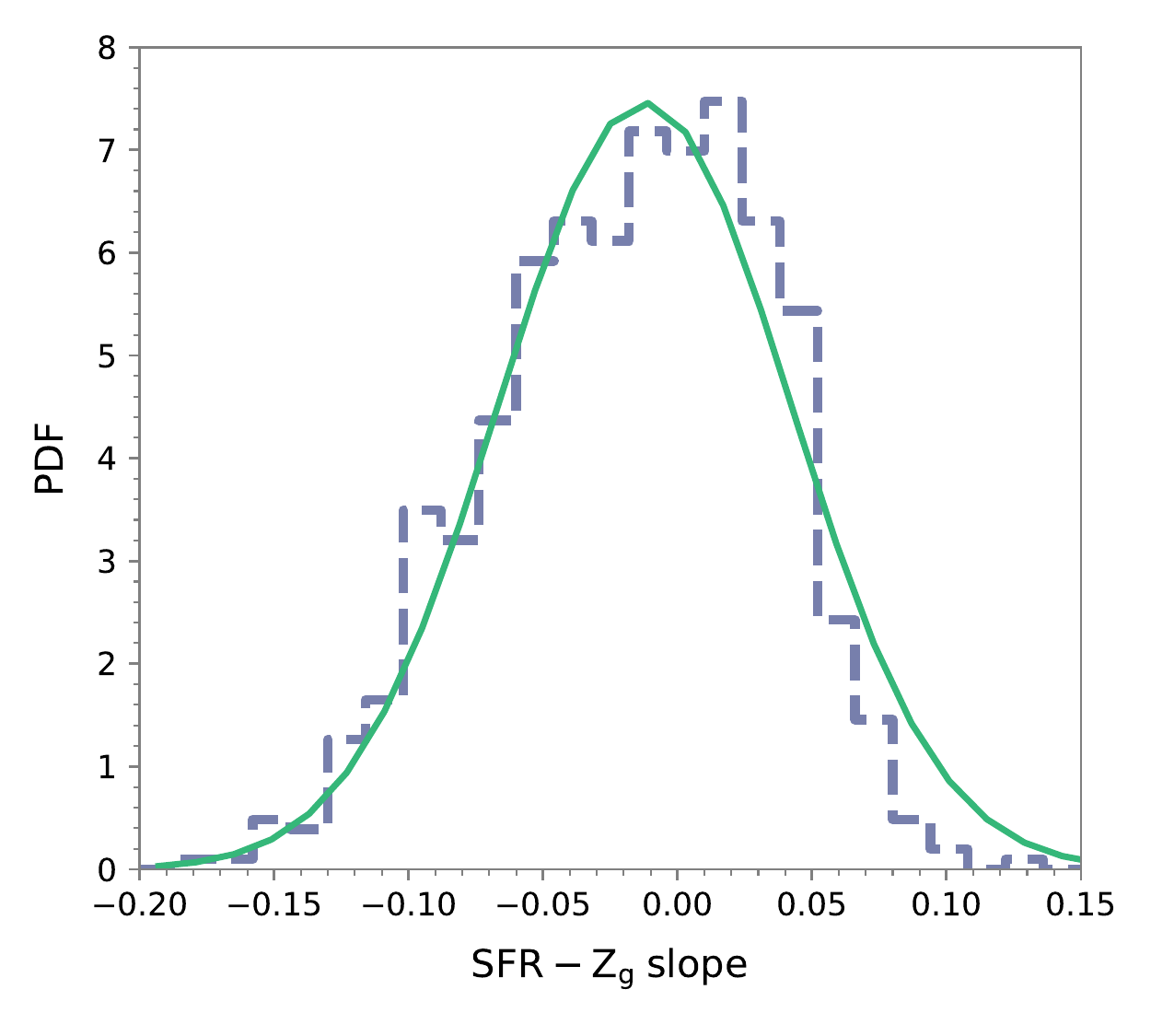}}
\caption{PDF of slopes of the correlations between SFR and $\rm Z_g$ residuals. Solid green line represents a Gaussian fit to the observed PDF (blue dashed line) sampled with the same bins.}
\label{fig3}
\end{center}
\end{figure}

The PDF of slopes of the linear fits for the entire sample is represented in Fig.~\ref{fig3} (blue dashed line). The measured slope values range from $-0.20$ to $0.15$, with the $60\%$ of the analyzed galaxies displaying a $\rm SFR-Z_g$ anti-correlation (i.e., negative slope) within the error bars. The remaining galaxies show no correlation (i.e., zero slope; $19\%$) or positive correlation (i.e., positive slope; $21\%$). We fit the PDF with a Gaussian function (green solid line) sampled using the same bins as the observed distribution. The peak of the distribution is found at $-0.012$, with a standard deviation of $0.054$. In order to estimate the contribution of the slope errors to the resulting width of the distribution we build 100 mock distributions of slopes in which each observed point is randomly shifted from the central value of $-0.012$ according to its measured error (assumed to follow a normal distribution). Gaussian fits to these mock distributions show that observational errors can only explain up to $\sim35 \%$ of the slope distribution width. This result suggests a physical origin for the large range of slope values displayed by the galaxies. In the next section we will investigate a possible dependence of the $\rm SFR-Z_g$ relation with different galaxy properties. 

\begin{figure}
\begin{center}
\resizebox{\hsize}{!}{\includegraphics{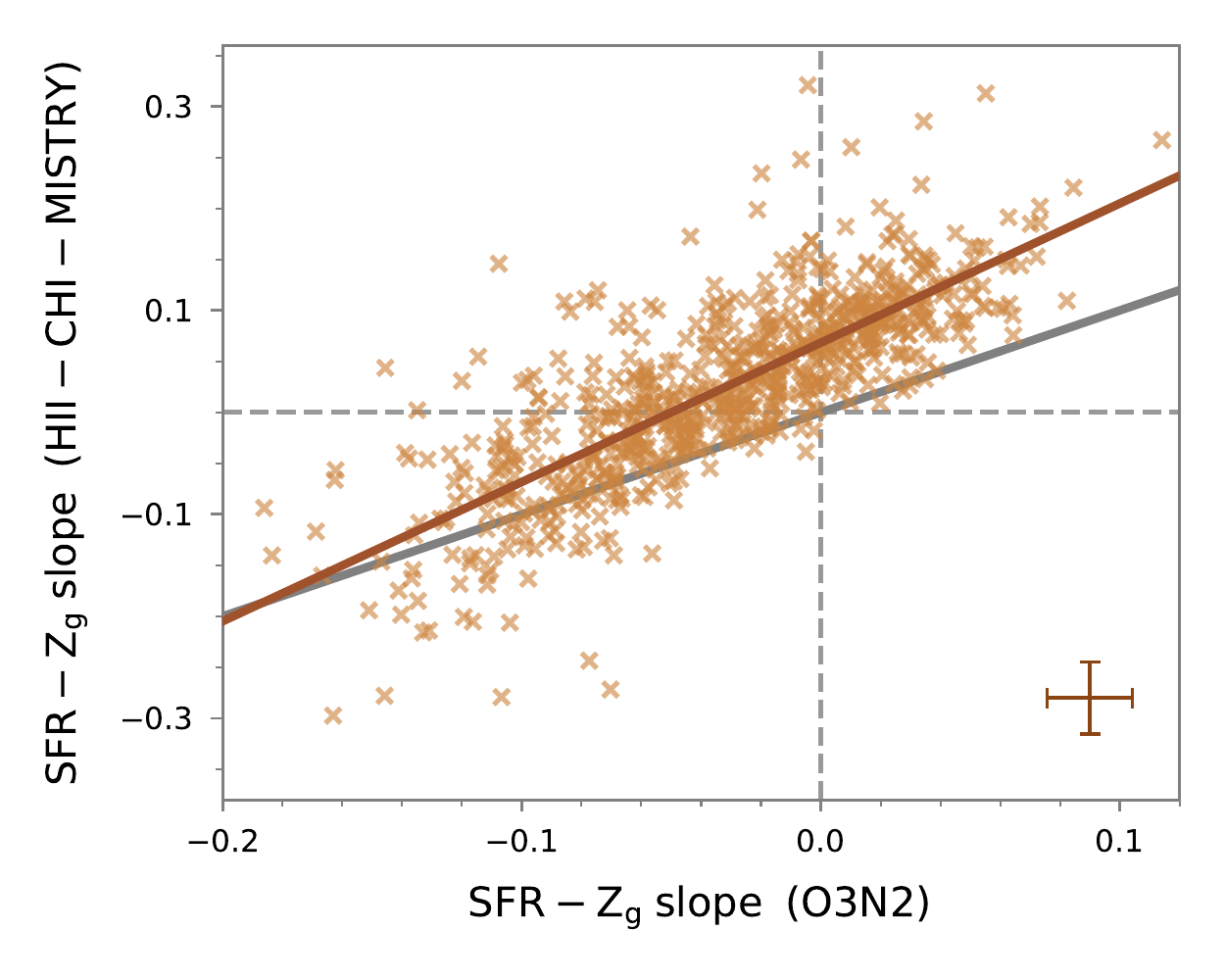}}
\caption{Comparison of the slopes of the local $\rm SFR-Z_g$ relation obtained using the M13 calibration for the O3N2 index (x-axis) and the HII-CHI-MISTRY code (y-axis) to determine $\rm Z_g$. The brown solid line corresponds to a linear regression of the individual points, and the grey solid line indicate the one-to-one relation. The dashed grey lines mark the zero value for the slopes. Error bars indicated in the bottom right corner correspond to the average slope errors.}
\label{fig8}
\end{center}
\end{figure}

Slopes of the local $\rm SFR-Z_g$ relation are also determined by applying the HII-CHI-MISTRY code to derive the gas-phase metallicities. The comparison of these values with the ones derived using the M13 calibration for the O3N2 index is represented in Fig.~\ref{fig8}. We can observe that the values corresponding to the HII-CHI-MISTRY calibration are higher than those obtained with the M13 calibration (brown solid line compared to the grey line representing the one-to-one relation). In this case, $35\%$ of the galaxies exhibit a $\rm SFR-Z_g$ anti-correlation, $55\%$ a positive correlation, and $10\%$ show signs of no correlation. This is not surprising, since discrepancies in the derivation of the oxygen abundances between multiple diagnostics and calibrations are broadly known \citep[for an extended discussion, see][]{kewley2008, lopezsanchez2012}. However, the correlation between both slopes is quite tight, allowing us to conclude that, although the actual measured values for the $\rm SFR-Z_g$ slopes differ, the qualitative results of the analysis on the dependence with different galaxy properties are equivalent for both calibrators. For the sake of clarity, below we only show the results based on the use of the M13 calibrator. The decision is based on the tighter local $\rm SFR-Z_g$ relations found for the galaxies with this calibrator. This is determined using the normalized root-mean-square error, NRMSE, preferred when comparing datasets with different scales:
\begin{equation}
{\rm NRMSE} = \frac{\sqrt{\sum_{i=1}^n \left( y_i - \hat y_i\right)^2/n}}{\left(Q_3 - Q_1\right)},
\end{equation}
where $Q_3 - Q_1$ is the difference between 75th and 25th percentiles, $y_i$ is the $i$-th observed abundance residual and $\hat y_i$ is the predicted value for the linear fit. The average NRMSE for M13 is $0.76$, against $0.81$ for HII-CHI-MISTRY. Despite this choice, we have reproduced the analysis using the HII-CHI-MISTRY code, with no significant differences between the obtained results. The reader can find this analysis in Appendix~\ref{sec:app2}. 

\section{Dependence with galaxy properties: The Random Forest approach}\label{sec:results2}
In this section we assess which galaxy properties may be determining in the local \mbox{$\rm SFR-Z_g$} relation by applying the statistical technique known as {\it Random Forests} (RF). It is a machine learning supervised algorithm proposed by \citet{breiman2001} which consists of an ensamble (combination) of decision trees. The purpose of a decision tree is to find the input features (the galaxy properties in our case) which contain the most information regarding the target feature (the derived slope of the $\rm SFR-Z_g$ relation) and creating a model to predict it defining a set of conditions on the values of the input features. When the features are categorical, we talk about classification decision trees (that lead to a RF classifier) and when they are numerical, regression decision trees (that lead to a RF regressor, as in our case). However decision trees have the disadvantages of presenting high variance (thus leading to an overfitting of the data) and being instable against small changes, that can lead to a very different tree structure. The RF algorithm allows to overcome these limitations by building multiple decision trees and averaging them based on the {\it bootstrap aggregation} technique (or bagging). This approach consists of (i) dividing the training sample into randomly-selected subsets with replacement and fit a decision tree to each of them, and (ii) considering a randomly-selected subset of features when splitting the trees. With this technique, the algorithm decreases the variance, reducing the chances of overfitting, as well as reduces the correlation between the trees. 

In this study we focus the analysis on eleven parameters which fully characterize the global properties of galaxies: total $M_\star$, average $\Sigma_\star$ at $R_{e}$, SFR, sSFR, morphological type, $g-r$ color, $\sigma_{cen}$, average LW stellar age and metallicity measured at $R_{e}$, oxygen abundance at $R_{e}$, and the slope of the oxygen abundance radial gradient. The goal of the RF is to assess which of these properties, if any, may be affecting the observed $\rm SFR-Z_g$ relation. The implementation of the algorithm was performed using the {\tt scikit-learn} package for Python \citep{pedregosa2011}. Next, we describe the basic steps followed to run the algorithm\footnote{We refer the reader to the {\tt scikit-learn} User Guide documentation for more details on the complete algorithm implementation, \url{https://scikit-learn.org/stable/modules/ensemble.html\#forest}}.

We first split the sample of 708 galaxies for which all the above mentioned properties and the slope of the $\rm SFR-Z_g$ relation are available into two subsets. The first subset, called training sample, corresponds to 2/3 of the galaxies (474) and it is used to create the model. For these galaxies, the selected set of eleven properties described above are provided to the algorithm to train the model (these are the {\it predictors}). The slope of the $\rm SFR-Z_g$ relation is considered the target feature aimed to be predicted by the RF model (this is the {\it solution}). The algorithm requires not only the predictors but also the solutions (called labels) to train the model. 

After training the model by building a set of decision trees, the RF allows to predict the slope of the $\rm SFR-Z_g$ relation of a new galaxy based on the values of its galaxy properties. However, once the model is trained, it is important to evaluate its performance on a new set of data. With this purpose, we employ the second subset of data, called the test sample, corresponding to the remaining 1/3 of the galaxies (234). Applying the model to the test sample, we obtain a set of predictions for the slope of the $\rm SFR-Z_g$ relation. The comparison of these predictions with the measured values provides an estimation of the precision of the model. 

Before training the model, in order to optimize the performance of the algorithm, it is recommended to constrain the model to make it simpler and less prone to overfitting. This is called regularisation. The amount of regularisation can be controlled by tuning the hyperparameters (HP), i.e. the parameters of the algorithm. In the case of a RF regressor, the main HPs include the number of trees in the forest ($n$), the number of randomly-selected features to consider in each split ($m$), the maximum depth of the trees ($max_{depth}$), the minimum number of samples required in each split ($min_{split}$), and the minimum number of samples that remain at the end of the different decision tree branches (stopping the algorithm from splitting the sample if its size is below this limit, $min_{leaf}$). To select the HPs that best suit the problem, a commonly used method is called K-Fold cross-validation (CV). This method consists in splitting the training sample into $K$ subsets, called folds, train the model on a different combination of $K-1$ folds and evaluate it on the remaining one. The performance measure reported is then the average of the values computed in the loop. Most authors suggest to perform K-Fold CV using $K = 5$ or $K = 10$ \citep{james2013}. Here we use $K=5$. We perform 100 iterations of the entire \mbox{5-Fold} CV process, each time using different values for the HPs. The selected HP values are the ones which achieve the highest average performance across the $K$-folds. In our case, it results in $n = 100$, $m = 6$, $max_{depth} = 120$, $min_{split} = 10$, and $min_{leaf} = 4$. 

\begin{figure}
\begin{center}
\resizebox{\hsize}{!}{\includegraphics{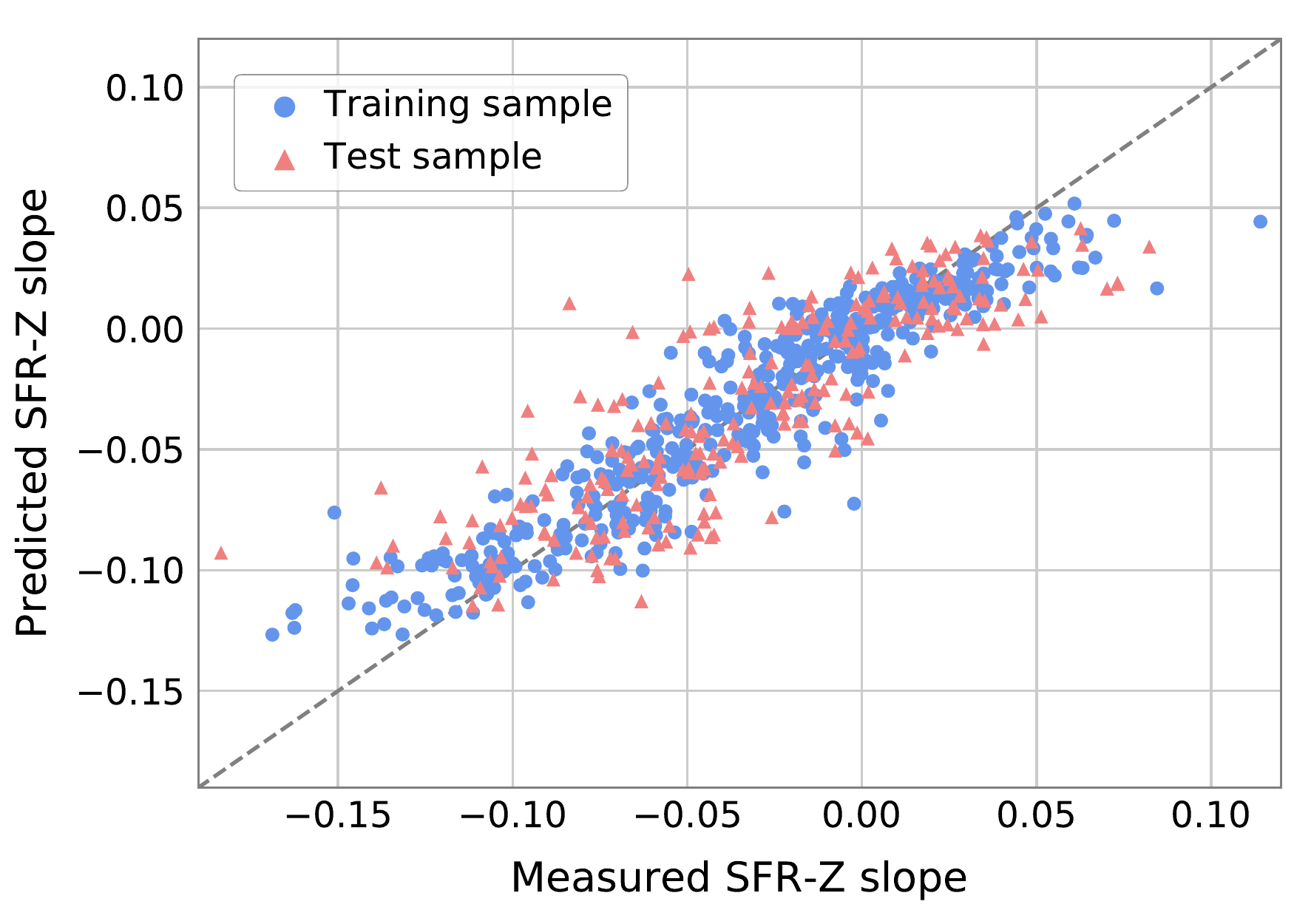}}
\caption{Slopes of the local $\rm SFR-Z_g$ relation predicted by the RF algorithm versus the measured values for the training (blue circles) and test (red triangles) samples. The dashed line indicates the one-to-one relation.}
\label{fig4}
\end{center}
\end{figure}

Finally, using the selected HPs, we train the model on the full training sample, and then evaluate on the test sample.  Figure~\ref{fig4} represents the predicted slopes of the $\rm SFR-Z_g$ relation by the RF algorithm against the measured values for both the training (blue circles) and test (red triangles) samples. We can see that the dispersion of the values around the one-to-one relation (dashed line) is similar for both samples (0.018 and 0.026, respectively), showing that the algorithm is able to predict with the same accuracy the slopes of galaxies that are not used to train the model. 

\begin{figure}
\begin{center}
\resizebox{\hsize}{!}{\includegraphics{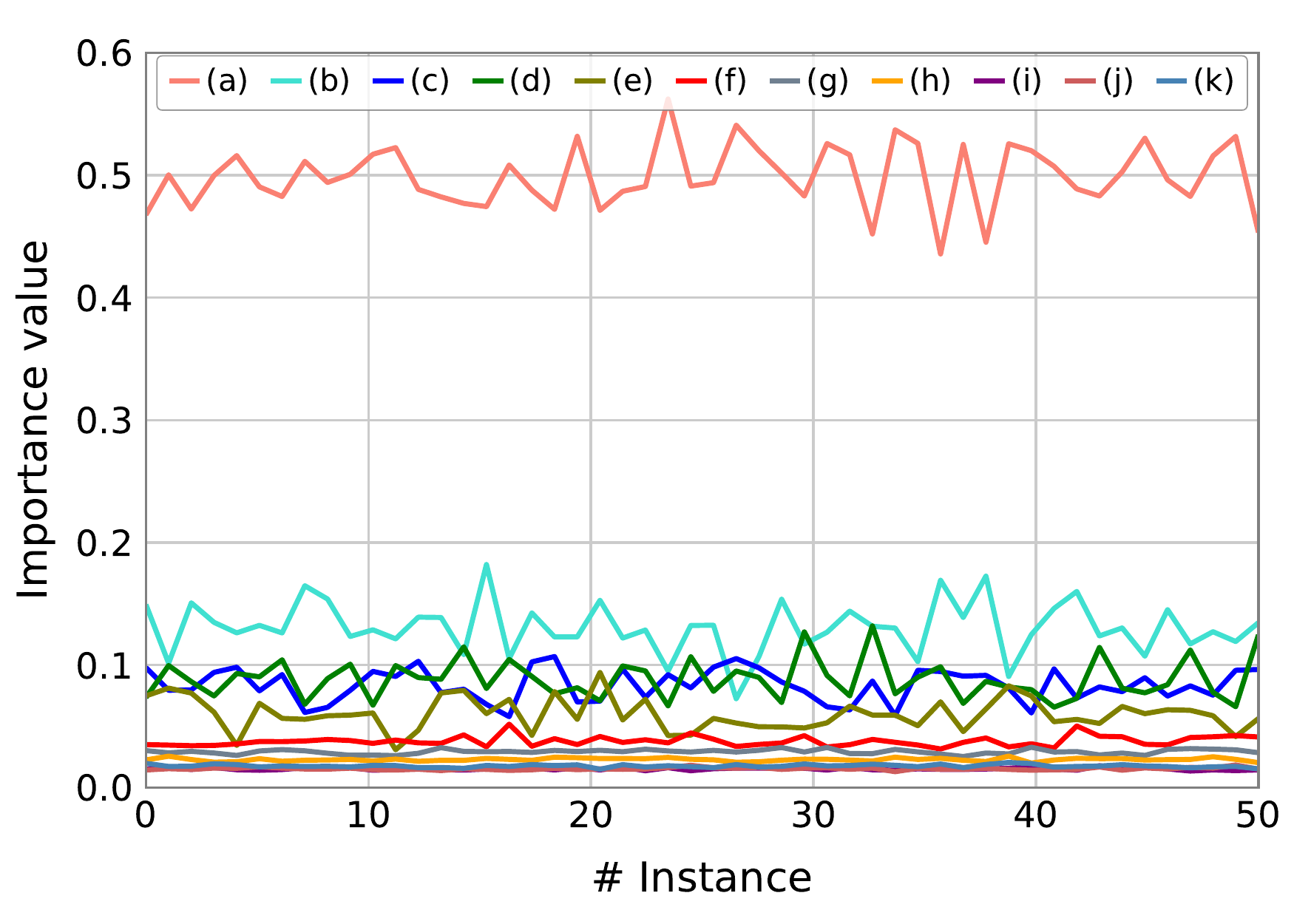}}
\caption{Relative {\it importance} of the input features of the RF in predicting the slope of the local $\rm SFR-Z_g$ relation for 50 runs of the algorithm. The different solid lines correspond to: (a) $\rm 12 + \log (O/H)$ at $R_{e}$, (b) $g-r$ color, (c) $M_\star$, (d) LW stellar age at $R_{e}$, (e) average $\Sigma_\star$ at $R_{e}$, (f) integrated SFR, (g) slope of the O/H radial gradient, (h) LW stellar Z at $R_{e}$, (i) morphological type, (j) $\sigma_{cen}$, and (k) sSFR.} 
\label{fig5}
\end{center}
\end{figure}

As mentioned before, some interesting information provided by the RF algorithm is the relative {\it importance} or contribution of each input feature in predicting the solution. These data are very useful, since they allow us to assess which of the analyzed properties of the galaxies have a high effect on the $\rm SFR-Z_g$ relation and which of them are completely irrelevant for determining the relation. In rough outlines, the feature importance is a measure of how effective the feature is at reducing variance when splitting the variables along the decision trees. The higher the value the more important the feature. In this analysis we obtain the following importance values for the predictors (in descending order): (a) oxygen abundance at $R_{e}$ (0.48), (b) $g-r$ color (0.12), (c) $M_\star$ (0.11), (d) LW stellar age at $R_{e}$ (0.09), (e) average $\Sigma_\star$ at $R_{e}$ (0.07), (f) SFR (0.03), (g) slope of the oxygen abundance radial gradient (0.03), (h) LW stellar Z at $R_{e}$ (0.02), (i) morphological type (0.02), (j) $\sigma_{cen}$ (0.02), and (k) sSFR (0.02). In order to test the stability of these values, we run the RF algorithm 50 times. Figure~\ref{fig5} shows the trend of the values for the 50 realizations. Although the actual values change slightly, the ranking of the features is almost always the same. The values for the importance clearly show that the global $\rm Z_g$ of a galaxy is the primary factor determining its local $\rm SFR-Z_g$ relation. Secondary factors ordered by decreasing importance are $g-r$ color, $M_\star$, average LW stellar age, and although in slightly less extent, average $\Sigma_\star$ at $R_{e}$. The remaining properties have proven to be of little relevance for the relation. 

\section{Polynomial regression analysis}\label{sec:results3}
The random forest approach has allowed us to identify the galaxy properties that might play a role at shaping the local $\rm SFR-Z_g$ relation. Among all the analyzed properties, the characteristic $\rm Z_g$ of the galaxies is the more determining factor, followed by $M_\star$, $g-r$ color, LW-age of the SP, and $\Sigma_\star$. The next step is finding a parametric model for the relationship between these properties and the slope of the $\rm SFR-Z_g$ relation. This is the basis of the polynomial regression presented in this section. 

\begin{figure*}
\begin{center}
\resizebox{\hsize}{!}{\includegraphics{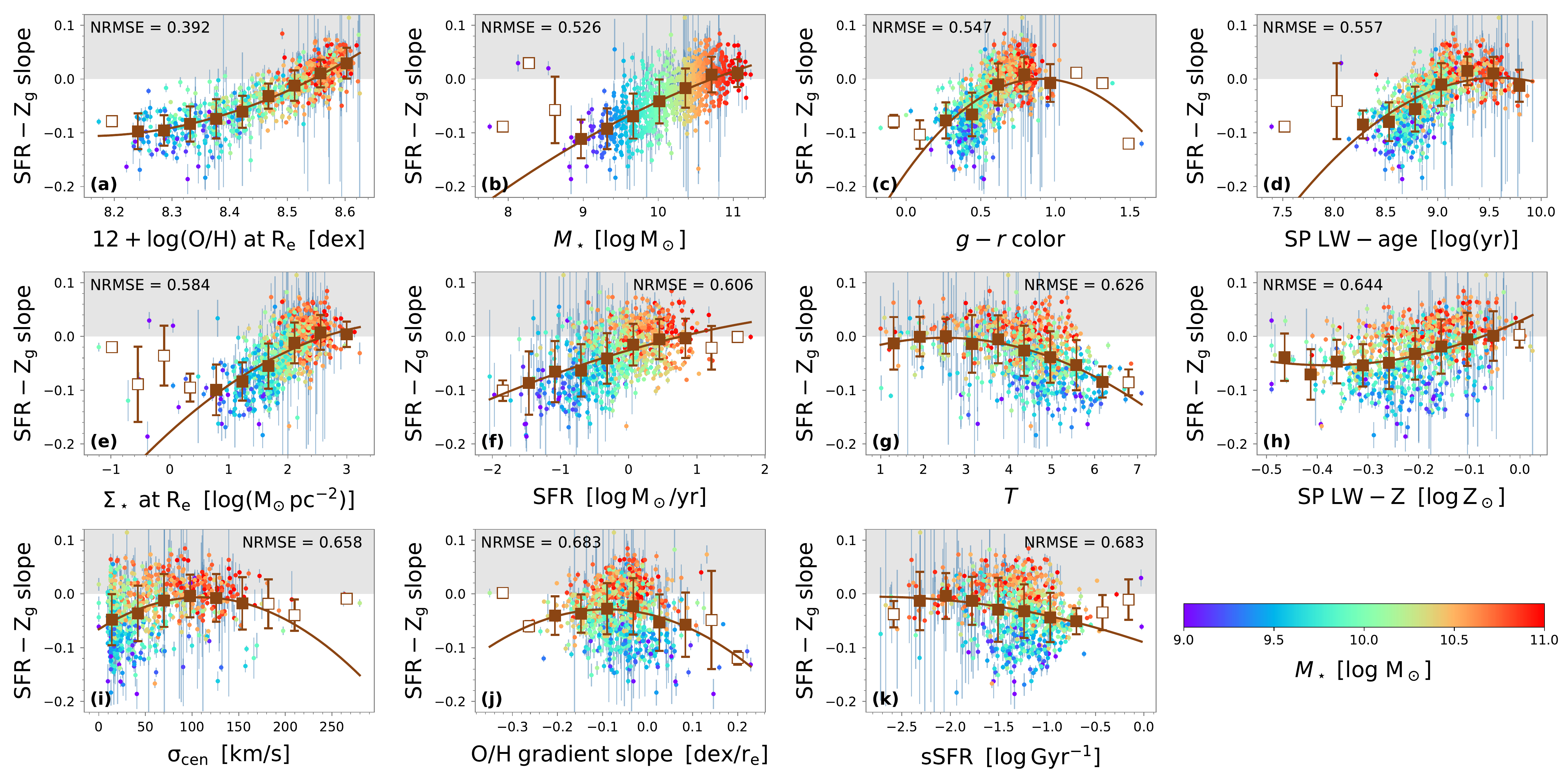}}
\caption{Slope of the local $\rm SFR-Z_g$ relation as a function of different galaxy properties, namely: (a) $\rm 12+\log (O/H)$ at $R_{e}$, (b) $M_\star$, (c) $g-r$ color, (d) LW stellar age at $R_{e}$, (e) average $\Sigma_\star$ at $R_{e}$, (f) integrated SFR, (g) morphological type, (h) LW stellar Z at $R_{e}$, (i) $\sigma_{cen}$, (j) slope of the $\rm O/H$ radial gradient, and (k) sSFR. Plots are color-coded according to $M_\star$. Brown squares represent the mean $\rm SFR-Z_g$ slope in the ten bins in which the galaxy parameters have been divided, with the error bars indicating the standard deviation of the slope within each bin. The averaged-binned values were fitted with a 2nd-order polynomial (brown solid lines), excluding the bins presenting less than 1\% of the total number of galaxies (empty squares). The normalized root-mean-square error (NRMSE) of the fits are given in each panel in ascending order from top left to bottom right.}
\label{fig6}
\end{center}
\end{figure*}

\begin{deluxetable*}{lCCC@{\hspace{1cm}}C}
\tablecaption{Coefficients and NRMSE of the 2nd-order polynomial fit to the relations between the local $\rm SFR-Z_g$ slope and the analyzed galaxy properties. \label{tab:coefs}}
\tablenum{1}
\tablehead{
\colhead{Galaxy property} &
\colhead{a$_0$} &
\colhead{a$_1$} & \colhead{a$_2$\hspace*{0.6cm}} & \colhead{NRMSE}
}
\startdata
$\rm Z_g$$^a$ at $R_{e}$ & +45.15295 & -11.10775 & +0.68153 & 0.392\\
$M_\star$ [log M$_\odot$] & -1.46183 & +0.22015 & -0.00781 & 0.526\\
$g-r$ color [mag] & -0.17349 & +0.38518 & -0.21335 & 0.547\\
LW stellar age at $R_{e}$ [log yr] & -5.50915 & +1.15029 & -0.06002 & 0.557\\
Average $\Sigma_\star$ at $R_{e}$ [log (M$_\odot$ pc$^{-2}$)] & -0.17756 & +0.09923 & -0.01201 & 0.584\\
SFR [log (M$_\odot$ yr$^{-1}$)] & -0.02564 & +0.03650 & -0.00407 & 0.606\\
Morphological type & -0.03876 & +0.02916 & -0.00585 & 0.626\\
LW stellar Z at $R_{e}$$^b$ & +0.02879 & +0.43084 & +0.56389 & 0.644\\
$\sigma_{cen}$ [km s$^{-1}$] & -6.5825\,10^{-2} & 1.0889\,10^{-3} & -4.9851\,10^{-6} & 0.658\\
Slope of the $\rm Z_g$$^a$ radial gradient [dex R$_e^{-1}$] & -0.03733 & -0.18969 & -1.03716 & 0.683\\
sSFR [log Gyr$^{-1}$] & -0.09028 & -0.06067 & -0.01083 & 0.683\\
\enddata
\tablenotetext{a}{With oxygen abundance as proxy, in the usual scale $\rm 12+\log\,(O/H)$.}
\tablenotetext{b}{In logarithmic scale and referred to the solar metallicity.}
\tablecomments{The polynomial fits have the form $p(x) = a_0 + a_1 \,x + a_2 \,x^2$.}
\end{deluxetable*}

Figure~\ref{fig6} shows the slope of the local $\rm SFR-Z_g$ relation as a function of the different galaxy properties provided as inputs to the RF algorithm. From top left to bottom right, the panels represent: (a) oxygen abundance at $R_{e}$, (b) $M_\star$, (c) $g-r$ color, (d) LW stellar age at $R_{e}$, (e) average $\Sigma_\star$ at $R_{e}$, (f) integrated SFR, (g) morphological type, (h) LW stellar Z at $R_{e}$, (i) $\sigma_{cen}$, (j) slope of the oxygen abundance radial gradient, and (k) sSFR. To see more clearly the general trend of the distributions, we represent (brown squares) the averaged $\rm SFR-Z_g$ slope values for ten bins in which the parameter range has been divided. The error bars indicate the standard deviation of the slope values within each bin. Focusing on the first six panels (a-f), that exhibit the clearest relations from all analyzed properties, we can see that the $\rm SFR-Z_g$ slope presents a positive correlation with all of them. The only anti-correlation appears for the morphological type, although the dispersion is high. It is worth mentioning that in the case of the characteristic $\rm Z_g$, the positive correlation seems to flatten at a slope value of around $-0.1$, suggesting the existence of a lower limit for the slope of the local $\rm SFR-Z_g$ relation for low metallicity galaxies (with $\rm 12 + \log (O/H)$ below $\sim 8.2-8.3$). Finally, the observed distributions are quite narrow until panel (e), corresponding to the galaxy properties for which the RF algorithm find significative contributions. From panel (f) on, the distributions widen until observing an almost absence of any relation for the last two parameters. 

In order to characterize the relations shown in Fig.~\ref{fig6} we fit a 2nd-order polynomial to all of them. This fit, represented by the brown solid lines, is performed on the averaged $\rm SFR-Z_g$ slope values of the bins (brown squares), excluding those presenting less than 1\% of the total number of galaxies (empty squares). The coefficients of the polynomials are provided in Table~\ref{tab:coefs}. In addition, the NRMSE of the fits are displayed in each panel and also provided in the same Table~\ref{tab:coefs}. Properties are ordered by the NRMSE in the figure (in ascending order). We can see that the characteristic $\rm Z_g$ presents the lowest NRMSE, supporting the RF-based result that this property is the fundamental factor in shaping the local $\rm SFR-Z_g$ relation. The other NRMSE values also go hand in hand with the RF conclusions, with the secondary parameters presenting the intermediate values and the ones with the lowest importance levels the highest NRMSE. 

Finally, once we have characterized the effect of the different galaxy properties on the slope of the local $\rm SFR-Z_g$ relation, one important issue we need to address is whether these relations present a secondary dependence with the remaining properties. The color-coding is a useful tool to investigate this issue. As an example, in Fig.~\ref{fig6} we color-code the individual points according to the stellar mass of the galaxies. In some cases we can see clear trends from left to right (panels a-f). These trends are the result of a dependence between the properties exhibited in these panels and the galaxy mass. For instance, we can see in the first panel that galaxies with the lowest metallicities are less massive systems. It is just the well-known MZR. However, a secondary dependence of the $\rm SFR-Z_g$ slope with any galaxy property should be reflected in the plots as a trend along the vertical axis, not along the horizontal direction. This trend is indeed observed in panels g-k, where we can see that the red symbols (massive galaxies) appear on the top of the distributions and the blue ones (less massive systems) show up on the bottom. This result is not surprising, since there is no dependence of the $\rm SFR-Z_g$ slope on these properties (supported by insignificant importance values in the RF), whereas a clear relation of this parameter with the galaxy mass has been found (both by the polynomial regression and the RF approach). 

\begin{figure}
\begin{center}
\resizebox{\hsize}{!}{\includegraphics{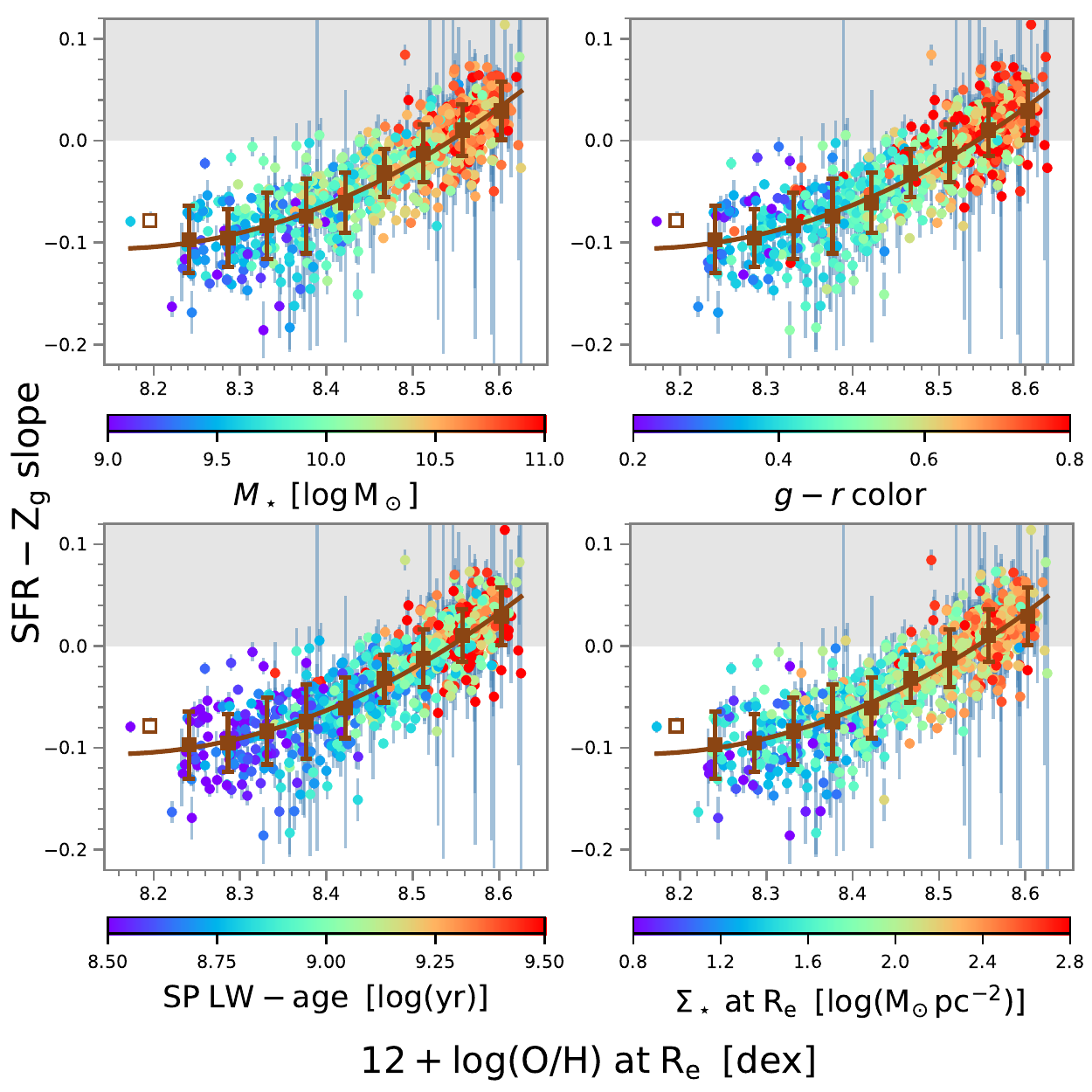}}
\caption{Slope of the local $\rm SFR-Z_g$ relation as a function of the gas-phase oxygen abundance at $R_{e}$. Plots are color-coded according to galaxy mass (top left), $g-r$ color (top right), LW-age of the stellar population at $R_{e}$ (bottom left), and average stellar mass density at $R_{e}$ (bottom right). See caption of Fig.~\ref{fig6} for more details.}
\label{fig7}
\end{center}
\end{figure}

In order to search for meaningful secondary dependences, we focus on the characteristic $\rm Z_g$ of the galaxies. The RF approach yielded the highest importance value for this galaxy property, which was supported by the smallest NRMSE obtained for the 2nd-order polynomial fit. If no secondary dependences are found in the relation between the $\rm SFR-Z_g$ slope and this galaxy property, we will be able to conclude that the global gas-phase metallicity of a galaxy is the fundamental factor shaping the local $\rm SFR-Z_g$ relation, with no additional influence of other properties. Figure~\ref{fig7} represents again the slope of the local $\rm SFR-Z_g$ relation as a function of $\rm Z_g$ at $R_{e}$. This time the distribution is color-coded according to the other four galaxy properties for which the RF algorithm found significant importance values: $M_\star$ (top left), $g-r$ color (top right), the average LW-age of the SP (bottom left), and the average $\Sigma_\star$ at $R_{e}$ (bottom right). As we can see, the vertical dispersion is not related to the color-code, that is, the dependence of the slope on the average gas metallicity does not exhibit any significant secondary dependence with respect to the rest of the galaxy properties. The local $\rm SFR-Z_g$ relation of a galaxy seems to be unambiguously characterized by its global gas metallicity. 

\section{Discussion and conclusions}\label{sec:discussion}
In this work we have studied a sample of 736 spiral galaxies in the Local Universe ($z<0.05$) using IFS data from the MaNGA survey. The 2D coverage of the data has allowed us to map local spatial variations of the gas-phase metallicity (using oxygen abundances as proxy) across entire galaxy discs. These variations have been obtained by subtracting the azimuthally averaged radial profile to the observed distribution \citep[radial profiles that show clear negative gradients in agreement with literature, see e.g.][among many others]{sanchez2014, ho2015, belfiore2017, sanchezmenguiano2016, sanchezmenguiano2018}. The resulting residual maps show the presence of significant chemical inhomogeneities across the discs (see left panels of Fig.~\ref{fig2}), with an amplitude up to $\rm \Delta \log Z_g \sim 0.2$ dex. This finding supports several works that brought to light recently the non-homogeneous nature of the chemical distribution of spiral galaxies \citep{sanchezmenguiano2016b, vogt2017, ho2017, ho2018, hwang2019}. 

Connecting these gas-phase chemical inhomogeneities with local variations of star formation rate ($\rm \Delta \log SFR$; derived in a similar way as for the metallicity), we find a local linear relation between both properties. The slope of this $\rm SFR-Z_g$ relation is negative for $\sim 60\%$ of the galaxies, and positive for $\sim 21\%$. The remaining $19\%$ are compatible with a zero slope. We note that these percentages are not volume corrected but derived specifically for the analyzed sample, and therefore, can not be extrapolated as characteristic of the Local Universe. Determining the values for a volume corrected sample goes beyond the scope of the paper because of the difficulties to quantify the selection function corresponding to the galaxies used in our study.

The finding of an anti-correlation between SFR and $\rm Z_g$ is not novel. Previous works arrived at this relation analysing a sample of star-forming dwarf galaxies in the Local Universe \citep{sanchezalmeida2013, sanchezalmeida2015, sanchezalmeida2018}. The large sample of galaxies available thanks to the MaNGA survey allows us to extend the analysis to a wider range not only of masses but also of other galaxy properties. This way, we are able to build a sample representative of galaxies in the Local Universe and put a larger frame to the picture. In this new frame, we find a non-negligible percentage of galaxies displaying a positive correlation between $\rm \Delta \log SFR$ and $\rm \Delta \log Z_g$ that was not present in previous works. The wide range of slopes obtained, that can not be explained by observational errors, evidence a physical origin for the diversified behaviour of the local $\rm SFR-Z_g$ relation.

In order to investigate its origin, we make use of a machine-learning technique known as Random Forests. This methodology has been extensively used in astronomy in the last decade with high levels of success, mainly in its classification form \citep[][among many others]{carliles2010, dubath2011, richards2011, bloom2012, brink2013, goldstein2015, liu2017, schanche2019}. Analysing a large set of galaxy properties, and building a model based on decision trees, the algorithm yields that the slope of the $\rm SFR-Z_g$ correlation is primarily determined by the average gas-phase metallicity of the galaxy. Galaxy mass, $g-r$ colors, stellar age, and mass density seem to play a less significant role. We note that this conclusion might be somewhat influenced by the galaxy sample used in the analysis. MaNGA galaxies were selected so that the number of galaxies per log mass bin is the same within a large range of stellar masses \citep[between $10^9$ and $10^{11}$ M$_\odot$; see][]{wake2017}. This selection is a good starting point for any exploratory study such as ours. Then we apply additional cuts in apparent size, inclination, and morphological type (Sect.~\ref{subsec:sample}), which modify the original distribution of masses and physical sizes as shown in Fig.~\ref{fig1}. Such cuts should not be determining provided the final sample reflects the full range of physical properties, and we have no reason to think that particular types of star-forming galaxies have been excluded from the final sample (see the wide range of galaxy properties considered in Fig.~\ref{fig6}, including those that seem to have little impact on the slope of the correlation).

The result on the galaxy properties setting the value of the slope is confirmed by the NRMSE obtained with a 2nd-order polynomial regression analysis, presenting the characteristic oxygen abundance with the lowest value. We find that the local $\rm SFR-Z_g$ slope varies almost linearly with the average gas-phase metallicity, being that the more metal-poor galaxies present the lowest slopes (i.e., the strongest $\rm SFR-Z_g$ anti-correlations). The dependence seems to flatten in the metal-poor regime, suggesting the existence of a single slope for low-metallicity galaxies (with $\rm 12 + \log (O/H)$ below $\sim 8.2-8.3$). However, this flattening is not observed when measuring the gas metallicities with HII-CHI-MISTRY (see Appendix~\ref{sec:app2} for details). Further analysis is needed in order to confirm or discard its existence. Finally, the analysis shows no secondary dependences of the relation between the $\rm SFR-Z_g$ slope and the average gas-phase metallicity on the rest of the galaxy properties. We can therefore conclude that the characteristic oxygen abundance of a galaxy unambiguously determine the shape of its local $\rm SFR-Z_g$ relation.

The performed analysis has shown to be robust in its main results, which we prove do not depend on the used methodology to derive the gas-phase metallicity. It is true that the O3N2 calibrator and HII-CHI-MISTRY yield different values for the slopes of the local $\rm SFR-Z_g$ relation, which results in a different percentage of observed anti-correlations ($60\%$ and $40\%$, respectively) and a different metallicity and galaxy mass value for the reversion of the slope from negative to positive ($\sim 8.5$ dex and $\rm 10.5 \,\log M_\odot$ against $8.4$ dex and $\rm 10.0 \,\log M_\odot$). However, the tight correlation between the slopes provided by both methods assure qualitatively equivalent results that lead to the global gas-phase metallicity of a galaxy as the primary factor determining the local relation between SFR and $\rm Z_g$. 

We also assess the effect of aperture bias in the determination of the stellar mass for the galaxies where the radial coverage of the instrument FoV only reaches out to $1.5 \,R_e$ (in contrast to galaxies where it reaches out to $2.5 \,R_e$). We repeat the analysis using masses from the NSA catalog, derived from the SDSS and GALEX photometric bands \citep{blanton2011}. The differences in the resulting importance values of the RF analysis are negligible, and although the NRMSE value of the polynomial fit for the galaxy mass is reduced from 0.526 to 0.479, it is still higher than the one obtained for the characteristic oxygen abundance. All of that reinforces that oxygen abundance is the most relevant factor affecting the local $\rm SFR-Z_g$ relation.

Various physical scenarios have been previously invoked to explain the observed anti-correlation between SFR and gas metallicity in dwarf galaxies \citep{sanchezalmeida2018}. One scenario is the local infall of metal-poor gas that mix with pre-existing gas and trigger star formation bursts, provided that both accreted and pre-existing gas in each star forming region have comparable masses. Our results suggest that this could be plausible for galaxies of low-intermediate mass ($\rm < 10-10.5 \log M_\odot$). On the contrary, for more massive systems the pre-existing gas mass may exceed the external gas mass, thus reversing the $\rm SFR-Z_g$ relation from having a negative slope to a positive one. Alternatively, the authors also proposed that the anti-correlation could be the effect of self-enrichment due to stellar winds and supernovae. However, for this to occur, the outflows driving the metals out of the star-forming regions should be very intense, with mass loading factors (i.e., the outflow rate in units of the SFR) larger than 10. Although large, it would be still consistent with some values found in local dwarf galaxies \citep[e.g.][]{martin1999, veilleux2005, olmogarcia2017}. 

Gas accretion of pristine gas has also been proposed by \citet{hwang2019} for the anomalously low metallicity regions found in a sample of late-type galaxies. These authors argue that the low-metallicity regions trace sites of recent accretion of gas that stimulates on-going star formation, in support of our findings. Furthermore, they find that the incidence rate of these regions is higher in lower mass galaxies. These results confirm our observations of a non-homogeneous chemical distribution of star forming regions driven by local infall of metal-poor gas that highly affect low-mass, metal-poor galaxies. 

Finally, some studies analysing the presence of variations in the gas chemistry of disc galaxies have pointed to self-enrichment associated with the spiral arms (and/or gas flows driven by the spiral pattern) as the cause of the presence of more metal-rich gas around these structures \citep{sanchezmenguiano2016b, vogt2017, ho2017, ho2018}. All analyzed galaxies correspond to quite massive systems ($\rm > 10.5 \log M_\odot$), which is in agreement with our findings of a positive $\rm SFR-Z_g$ correlation at this mass range as a result of localized metal recycling by pre-existing gas. 

All these results evidence that the time-scales of mixing processes taking place in spiral galaxies are large enough as to allow the durability of chemical variations for long periods of time. In addition, the persistence of external inhomogeneous metal-poor gas infall could be an alternative driver of chemical differences for low-intermediate mass galaxies.

\vspace*{0.8cm}
In summary, this work shows the existence of a linear relation between log SFR and $\rm \log \,Z_g$ at local scales (after correcting for radial trends), with the slope determined mainly by the average gas-phase metallicity of galaxies. We confirm previous studies finding an anti-correlation for dwarf galaxies, which reverses for more metal-rich (and massive) systems. As a plausible explanation, we propose a scenario in which external gas accretion fuels star-formation in metal-poor galaxies, whereas the gas comes from previous star formation episodes in metal-rich systems. Numerical simulations emerge as key in confirming this framework and provide an answer to the level of contribution of gas accretion to galaxy formation and chemical evolution.

\newpage
\acknowledgements
We acknowledge financial support from the Spanish Ministerio de Econom\'ia y Competitividad (MINECO) via {\it Estallidos} grant AYA2016-79724-C4-2-P. MF gratefully acknowledges the financial support of the ``Funda\c{c}\~{a}o para a Ci\^{e}ncias e Tecnologia'' (FCT - Portugal), through the grant SFRH/BPD/107801/2015.

We thank the anonymous referee for their suggestions that allowed us to improve the paper. Thanks are also due to Dalya Baron and Marc Huertas-Company for help with the Random Forest tools and analysis.

This project makes use of the MaNGA-Pipe3D dataproducts. We thank the IA-UNAM MaNGA team for creating it, and the ConaCyt-180125 project for supporting them.

Funding for the Sloan Digital Sky Survey IV has been provided by the Alfred P. Sloan Foundation, the U.S. Department of Energy Office of Science, and the Participating Institutions. SDSS acknowledges support and resources from the Center for High-Performance Computing at the University of Utah. The SDSS web site is \url{www.sdss.org}.

SDSS is managed by the Astrophysical Research Consortium for the Participating Institutions of the SDSS Collaboration including the Brazilian Participation Group, the Carnegie Institution for Science, Carnegie Mellon University, the Chilean Participation Group, the French Participation Group, Harvard-Smithsonian Center for Astrophysics, Instituto de Astrofísica de Canarias, The Johns Hopkins University, Kavli Institute for the Physics and Mathematics of the Universe (IPMU) / University of Tokyo, the Korean Participation Group, Lawrence Berkeley National Laboratory, Leibniz Institut f\"ur Astrophysik Potsdam (AIP), Max-Planck-Institut f\"ur Astronomie (MPIA Heidelberg), Max-Planck-Institut f\"ur Astrophysik (MPA Garching), Max-Planck-Institut f\"ur Extraterrestrische Physik (MPE), National Astronomical Observatories of China, New Mexico State University, New York University, University of Notre Dame, Observat\'orio Nacional / MCTI, The Ohio State University, Pennsylvania State University, Shanghai Astronomical Observatory, United Kingdom Participation Group, Universidad Nacional Aut\'onoma de México, University of Arizona, University of Colorado Boulder, University of Oxford, University of Portsmouth, University of Utah, University of Virginia, University of Washington, University of Wisconsin, Vanderbilt University, and Yale University.

\appendix

\section{Galaxy properties}\label{sec:app1}
In this section we present Table~\ref{tab:app1} containing information on the galaxy properties analyzed along the article together with the determined slopes for the local $\rm SFR-Z_g$ relation. For information on the derivation of these parameters we refer the reader to Sect.~\ref{sec:analysis} and \ref{sec:results1}.  

\vspace*{0.4cm}

\section{Characterisation of the local $\rm SFR-Z_{\lowercase{g}}$ relation using HII-CHI-MISTRY}\label{sec:app2}
\begin{figure}[!h]
\plottwo{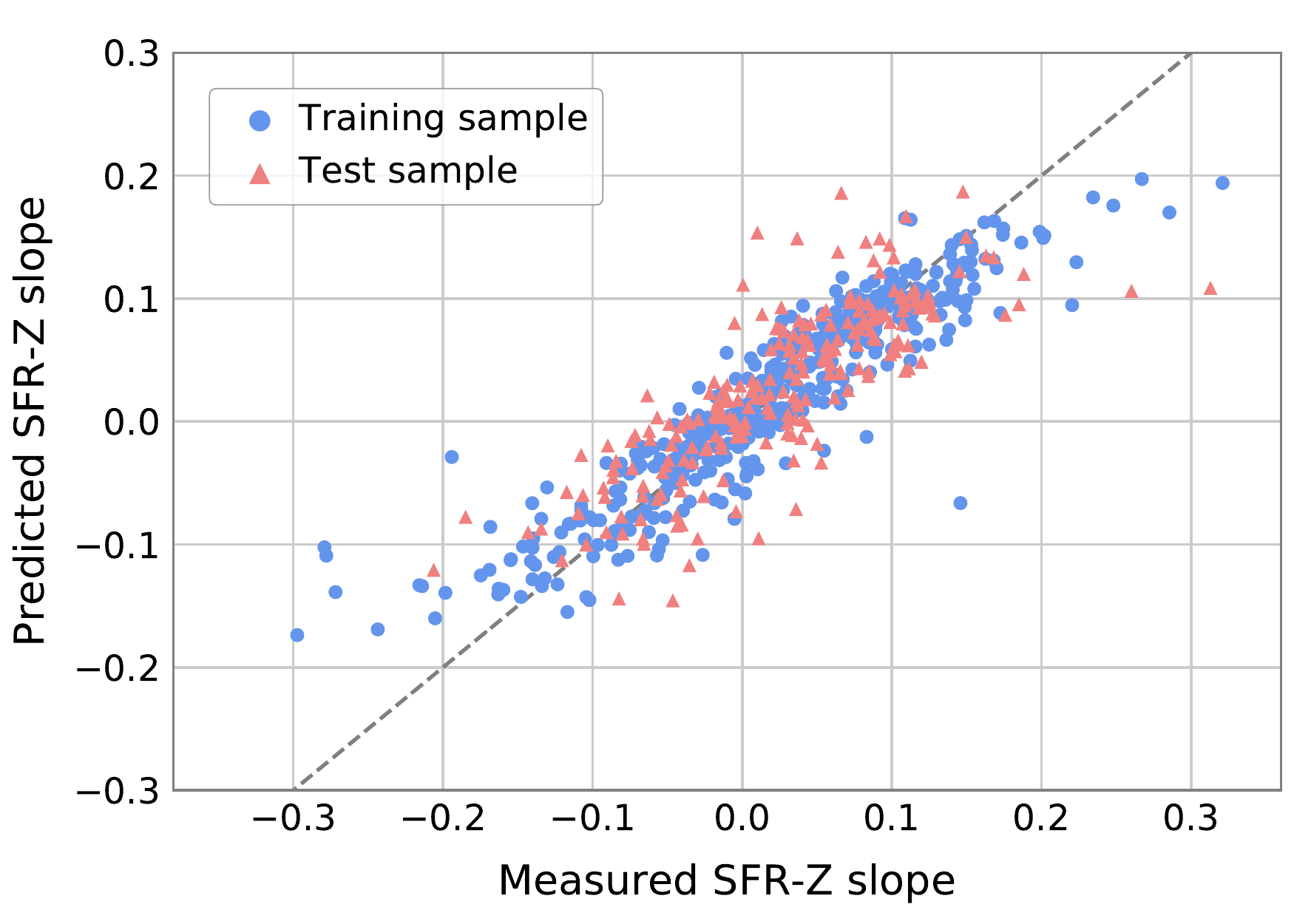}{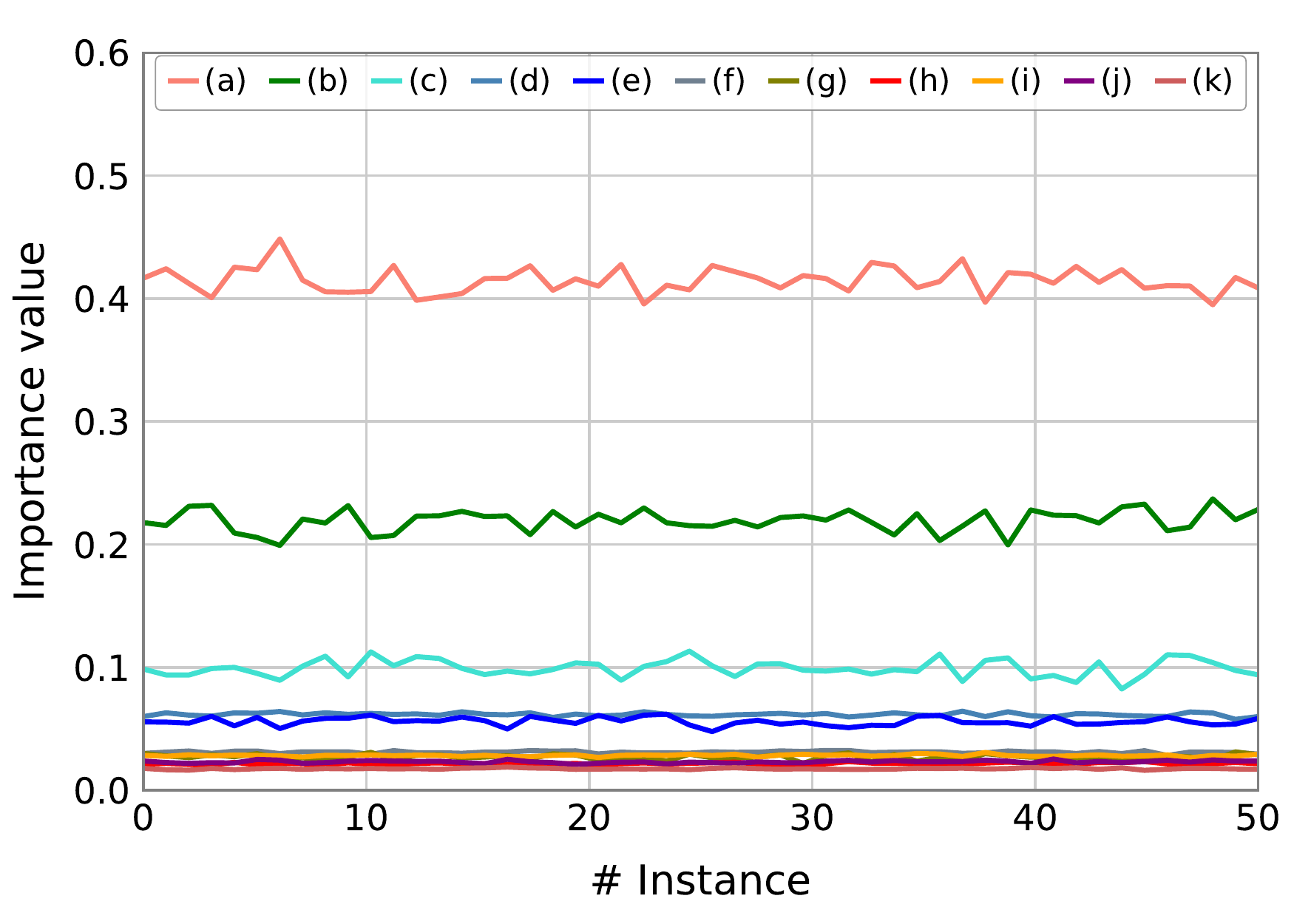}
\caption{RF algorithm with HII-CHI-MISTRY: prediction of slopes of the local $\rm SFR-Z_g$ relation ({\it left}) and relative importance of the input features for 50 runs of the algorithm ({\it right}). The different solid lines correspond to: (a) $\rm 12+\log (O/H)$ value at $R_{e}$, (b) LW stellar age at $R_{e}$, (c) $g-r$ color, (d) sSFR, (e) slope of the $\rm O/H$ radial gradient, (f) integrated SFR, (g) LW stellar Z at $R_{e}$, (h) $M_\star$, (i) morphological type, (j) $\sigma_{cen}$, and (k) average $\Sigma_\star$ at $R_{e}$. See captions of Fig.~\ref{fig4} and \ref{fig5} for details.}
\label{fig10}
\end{figure}

In this appendix we assess the effect of using the HII-CHI-MISTRY calibrator on the results presented in the paper. With such purpose, we replicate the RF analysis to study the dependence of the local $\rm SFR-Z_g$ relation on the different galaxy properties. For the regularisation of the model, the performed 5-Fold CV procedure yields the following HPs: $n = 600$, $m = 6$, $max_{depth} = 40$, $min_{split} = 6$, and $min_{leaf} = 5$. With these HPs, the RF algorithm provides the predicted slopes of the $\rm SFR-Z_g$ relation shown in the left panel of Fig.~\ref{fig10}. Both the training (blue circles) and the test (red triangles) samples are distributed around the one-to-one relation (dashed line) covering the same parameter space with a similar dispersion (0.036 and 0.045 respectively). Regarding the relative importance of the input features, we obtain the following values (in descending order): (a) oxygen abundance at $R_{e}$ (0.40), (b) LW stellar age at $R_{e}$ (0.24), (c) $g-r$ color (0.10), (d) sSFR (0.06), (e) galaxy mass (0.05), (f) LW stellar metallicity at $R_{e}$ (0.03), (g) slope of the oxygen abundance radial gradient (0.03), (h) average stellar mass density at $R_{e}$ (0.03), (i) integrated SFR (0.02), (j) morphological type (0.02), and (k) $\sigma_{cen}$ (0.02). The right panel of Fig.~\ref{fig10} displays the trend of the values for 50 realizations, showing the high stability of the feature importances. In agreement with the results obtained with the O3N2 calibrator, the gas-phase metallicity remains as the primary factor determining its local $\rm SFR-Z_g$ relation. In this case, average LW stellar age, $g-r$ color, sSFR, and stellar mass would be the next factors by importance order. The remaining properties result of little relevance for the relation. 

\begin{figure}
\begin{center}
\resizebox{\hsize}{!}{\includegraphics{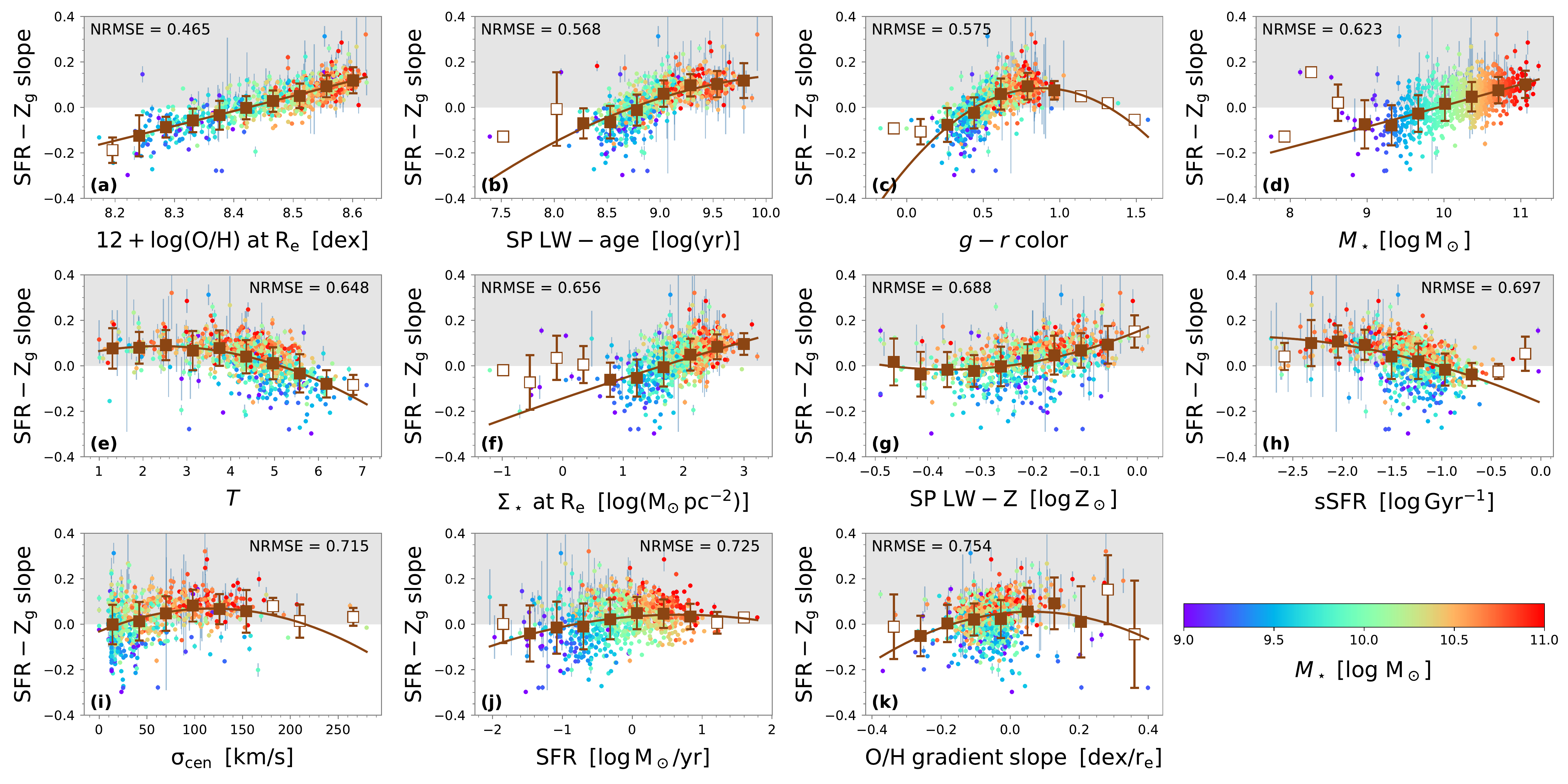}}
\caption{Slope of the local $\rm SFR-Z_g$ relation derived with the HII-CHI-MISTRY calibrator as a function of different galaxy properties, namely: (a) $\rm 12+\log (O/H)$ value at $R_{e}$, (b) LW stellar age at $R_{e}$, (c) $g-r$ color, (d) $M_\star$, (e) morphological type, (f) average $\Sigma_\star$ at $R_{e}$, (g) LW stellar Z at $R_{e}$, (h) sSFR, (i) $\sigma_{cen}$, (j) integrated SFR, and (k) slope of the $\rm O/H$ radial gradient. See caption of Fig.~\ref{fig6} for more details.}
\label{fig11}
\end{center}
\end{figure}

\begin{figure}
\begin{center}
\resizebox{\hsize}{!}{\includegraphics{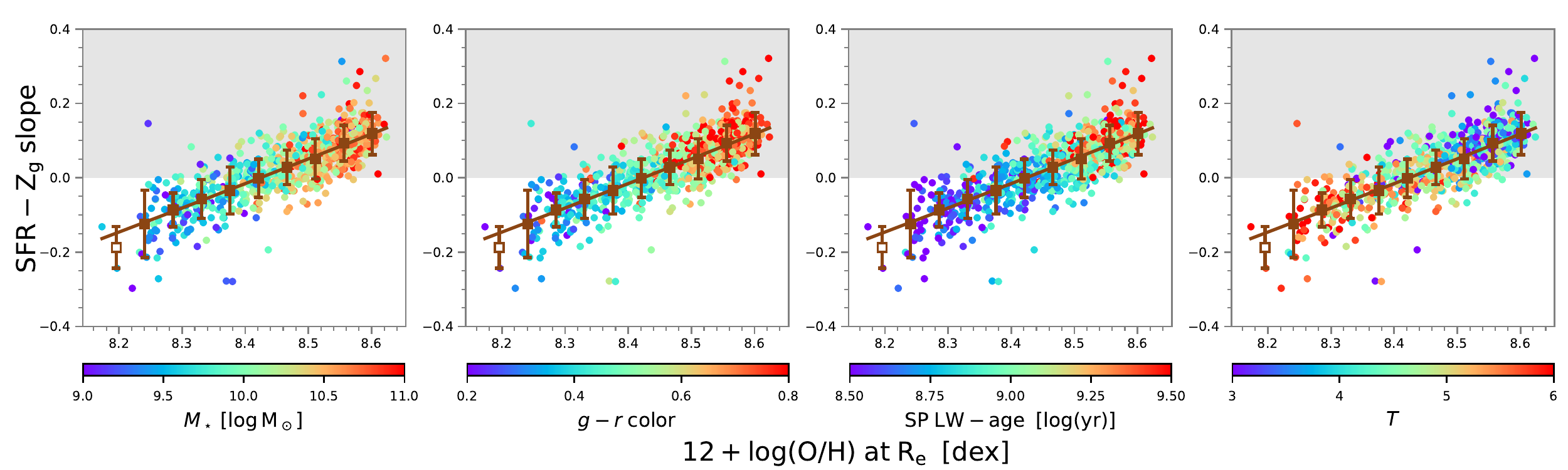}}
\caption{Slope of the local $\rm SFR-Z_g$ relation derived with the HII-CHI-MISTRY as a function of $\rm 12+\log (O/H)$ at $R_{e}$. Plots are color-coded according to $M_\star$ (left), $g-r$ color (middle left), LW stellar age at $R_{e}$ (middle right), and morphological type (right). See caption of Fig.~\ref{fig6} for more details.}
\label{fig12}
\end{center}
\end{figure}

We replicate the polynomial regression analysis for the dependence of the $\rm SFR-Z_g$ slope with the galaxy properties, whose outcome we present in Fig.~\ref{fig11}. The shapes of the relations are the same than those determined using the O3N2 calibrator, although the specific NRMSE values of the fits change (but still supporting the conclusions of the RF). We see again that the characteristic oxygen abundance presents the lowest NRMSE, significantly below the rest of values. In this case, the flattening of the relation at the low metallicity regime found with O3N2 is not observed. Finally, the HII-CHI-MISTRY calibrator also yields no secondary dependence of the relation between the $\rm SFR-Z_g$ slope and the average gas metallicity with the rest of galaxy properties (see Fig.~\ref{fig12}). We can therefore confirm the conclusion of the local $\rm SFR-Z_g$ relation of a galaxy being unambiguously characterized by its characteristic oxygen abundance, independently of the employed calibrator to derive the gas-phase metallicity.

\vspace*{0.3cm}

\begin{longrotatetable}
\begin{deluxetable*}{lCCCCCCCCCCCC}
\tablecaption{Galaxy properties and derived slopes for the local $\rm SFR-Z_g$ relation. \label{tab:app1}}
\tabletypesize{\scriptsize}
\tablenum{2}
\tablehead{
\colhead{Galaxy ID} & \colhead{[12 + log(O/H)]$_{Re}$} & \colhead{$\rm log \,M_{*}$} & 
\colhead{$g-r$} & \colhead{LW-age} & \colhead{$\rm log \,\Sigma_{*}$} & 
\colhead{log SFR} & \colhead{T} & \colhead{LW-Z} & \colhead{$\rm \sigma_{cen}$} & 
\colhead{$\rm [O/H]_{slope}$} & \colhead{sSFR} & \colhead{$\rm SFR-Z_g$ slope} \\
\colhead{} & \colhead{[dex]} & \colhead{[M$_\odot$]} & 
\colhead{} & \colhead{[log yr]} & \colhead{[M$_\odot$ pc$^{-2}$]} & 
\colhead{[M$_\odot$ yr$^{-1}$]} & \colhead{} & \colhead{[log Z$_\odot$]} & \colhead{[km s$^{-1}$]} & 
\colhead{[dex R$_e^{-1}$]} & \colhead{[log Gyr$^{-1}$]} & \colhead{}
}
\colnumbers
\startdata
9487-12701  &  8.50  &  10.43  &  0.67  &  9.00  &  1.66  &  0.08  &  5.9  &  -0.05  &  61.08  &  -0.20  &  -1.36  &  -0.021 \pm 0.006  \\ 
8485-12701  &  8.29  &  9.57  &  0.34  &  8.73  &  1.79  &  -0.52  &  6.0  &  -0.33  &  22.92  &  -0.06  &  -1.09  &  -0.105 \pm 0.005  \\ 
8134-12701  &  8.39  &  9.62  &  0.48  &  8.72  &  1.28  &  -0.75  &  4.2  &  -0.26  &  17.36  &  -0.17  &  -1.38  &  -0.112 \pm 0.007  \\ 
8552-12701  &  8.47  &  10.66  &  0.89  &  9.27  &  2.70  &  0.20  &  4.7  &  -0.24  &  114.16  &  -0.02  &  -1.46  &  +0.014 \pm 0.008  \\ 
9492-12701  &  8.47  &  10.55  &  0.58  &  8.65  &  1.80  &  0.75  &  5.5  &  -0.25  &  46.83  &  -0.19  &  -0.80  &  -0.060 \pm 0.003  \\ 
8985-12701  &  8.36  &  9.80  &  0.37  &  8.88  &  1.35  &  -0.38  &  5.1  &  -0.42  &  19.69  &  0.00  &  -1.18  &  -0.068 \pm 0.002  \\ 
8312-12701  &  \mbox{---}   &  9.52  &  0.35  &  \mbox{---}   &  \mbox{---}   &  -0.40  &  4.8  &  \mbox{---}   &  9.71  &  \mbox{---}   &  -0.92  &  -0.092 \pm 0.007  \\ 
9505-12701  &  8.56  &  11.07  &  0.83  &  9.47  &  2.61  &  0.63  &  4.3  &  -0.16  &  126.82  &  -0.06  &  -1.44  &  +0.007 \pm 0.016  \\ 
7990-12701  &  8.54  &  10.19  &  0.52  &  8.96  &  1.75  &  -0.19  &  5.2  &  -0.19  &  12.91  &  -0.05  &  -1.38  &  +0.035 \pm 0.011  \\ 
8243-12701  &  8.54  &  11.16  &  0.83  &  9.20  &  2.41  &  0.99  &  5.3  &  -0.23  &  119.16  &  -0.12  &  -1.17  &  +0.010 \pm 0.006  \\ 
8338-12701  &  8.31  &  9.96  &  0.24  &  8.65  &  1.70  &  -0.07  &  6.1  &  -0.30  &  15.19  &  -0.08  &  -1.03  &  -0.074 \pm 0.003  \\ 
9510-12701  &  8.49  &  10.23  &  0.50  &  8.88  &  1.78  &  0.00  &  5.3  &  -0.23  &  12.99  &  -0.13  &  -1.22  &  -0.032 \pm 0.004  \\ 
8323-12701  &  8.30  &  9.99  &  0.35  &  8.53  &  1.65  &  0.14  &  6.7  &  -0.33  &  12.99  &  -0.02  &  -0.85  &  -0.082 \pm 0.004  \\ 
8993-12701  &  8.57  &  10.60  &  0.84  &  9.02  &  \mbox{---}   &  -0.09  &  4.9  &  -0.23  &  90.91  &  0.02  &  -1.68  &  +0.046 \pm 0.009  \\ 
7991-12701  &  8.50  &  10.65  &  0.65  &  8.82  &  2.39  &  0.85  &  5.1  &  -0.32  &  63.42  &  -0.16  &  -0.80  &  -0.081 \pm 0.004  \\ 
9870-12701  &  8.35  &  9.04  &  0.45  &  8.62  &  0.88  &  -1.51  &  3.3  &  -0.24  &  9.81  &  -0.03  &  -1.55  &  -0.16 \pm 0.03  \\ 
8274-12701  &  8.51  &  9.72  &  0.58  &  8.92  &  2.40  &  -0.42  &  4.2  &  -0.20  &  9.75  &  0.00  &  -1.14  &  -0.016 \pm 0.009  \\ 
8948-12701  &  8.43  &  10.09  &  0.51  &  8.75  &  2.01  &  0.01  &  5.3  &  -0.33  &  21.27  &  -0.12  &  -1.09  &  -0.058 \pm 0.005  \\ 
8716-12701  &  8.37  &  8.84  &  0.46  &  8.96  &  0.98  &  -1.97  &  2.9  &  -0.09  &  17.72  &  -0.04  &  -1.81  &  -0.11 \pm 0.06  \\ 
8718-12701  &  8.52  &  10.64  &  0.70  &  9.17  &  2.13  &  0.26  &  4.1  &  0.01  &  70.28  &  -0.00  &  -1.38  &  +0.001 \pm 0.015  \\ 
8567-12701  &  \mbox{---}   &  10.00  &  0.50  &  8.83  &  \mbox{---}   &  -0.30  &  4.0  &  -0.13  &  13.14  &  \mbox{---}   &  -1.31  &  +0.051 \pm 0.004  \\ 
8155-12701  &  8.54  &  10.64  &  0.63  &  9.04  &  2.07  &  0.42  &  3.8  &  -0.21  &  41.50  &  -0.17  &  -1.22  &  -0.001 \pm 0.005  \\ 
8329-12701  &  8.59  &  10.84  &  0.67  &  9.12  &  2.10  &  0.59  &  4.9  &  -0.17  &  69.49  &  -0.09  &  -1.25  &  +0.040 \pm 0.006  \\ 
8080-12701  &  8.48  &  10.59  &  0.50  &  8.72  &  1.70  &  0.56  &  5.0  &  -0.13  &  20.43  &  -0.20  &  -1.03  &  -0.049 \pm 0.005  \\ 
7964-12701  &  8.29  &  9.51  &  0.37  &  8.57  &  1.07  &  -0.61  &  5.9  &  -0.17  &  16.74  &  -0.08  &  -1.12  &  -0.141 \pm 0.008  \\ 
8458-12701  &  8.52  &  10.43  &  0.70  &  8.96  &  2.33  &  0.57  &  3.0  &  -0.19  &  135.73  &  0.01  &  -0.86  &  +0.02 \pm 0.03  \\ 
7960-12701  &  8.59  &  11.01  &  0.67  &  9.14  &  2.39  &  0.49  &  4.9  &  -0.36  &  16.22  &  0.00  &  -1.52  &  +0.017 \pm 0.004  \\ 
8715-12701  &  8.59  &  10.63  &  0.81  &  9.52  &  2.30  &  -0.27  &  4.7  &  -0.26  &  64.48  &  -0.05  &  -1.90  &  +0.036 \pm 0.013  \\ 
9037-12701  &  8.29  &  9.70  &  0.38  &  8.73  &  1.79  &  -0.82  &  5.2  &  -0.39  &  158.42  &  -0.06  &  -1.52  &  -0.102 \pm 0.006  \\ 
8313-12701  &  8.51  &  10.78  &  0.63  &  8.89  &  2.13  &  0.80  &  5.0  &  -0.20  &  57.58  &  -0.08  &  -0.98  &  -0.007 \pm 0.003  \\ 
8241-12701  &  8.34  &  9.75  &  0.44  &  8.62  &  1.56  &  -0.86  &  5.1  &  -0.23  &  9.79  &  -0.06  &  -1.61  &  -0.071 \pm 0.010  \\ 
8592-12701  &  8.26  &  9.56  &  0.33  &  8.37  &  1.61  &  -0.44  &  5.6  &  -0.30  &  25.36  &  -0.04  &  -1.00  &  -0.070 \pm 0.010  \\ 
9872-12701  &  8.32  &  9.56  &  0.39  &  8.61  &  1.29  &  -0.70  &  5.6  &  -0.27  &  13.05  &  -0.07  &  -1.26  &  -0.076 \pm 0.004  \\ 
8952-12701  &  8.62  &  10.17  &  0.77  &  9.15  &  1.98  &  -0.43  &  4.2  &  -0.03  &  76.27  &  0.06  &  -1.59  &  +0.08 \pm 0.06  \\ 
8450-12701  &  8.55  &  10.12  &  0.54  &  9.01  &  1.96  &  -0.23  &  2.9  &  -0.24  &  27.34  &  -0.10  &  -1.34  &  +0.032 \pm 0.013  \\ 
8153-12701  &  8.40  &  9.85  &  0.46  &  8.65  &  1.28  &  -0.51  &  5.2  &  -0.06  &  24.49  &  -0.08  &  -1.36  &  -0.096 \pm 0.014  \\ 
8591-12701  &  8.36  &  9.74  &  0.42  &  8.78  &  1.20  &  -0.22  &  5.0  &  -0.21  &  12.77  &  -0.11  &  -0.96  &  -0.08 \pm 0.02  \\ 
8439-12701  &  8.32  &  9.32  &  0.45  &  8.62  &  1.29  &  -1.23  &  5.1  &  -0.29  &  13.08  &  -0.00  &  -1.55  &  -0.106 \pm 0.011  \\ 
8444-12701  &  8.36  &  9.43  &  0.51  &  8.85  &  1.57  &  -1.51  &  4.6  &  -0.29  &  13.08  &  0.02  &  -1.93  &  -0.18 \pm 0.02  \\ 
9048-12701  &  8.38  &  10.01  &  0.45  &  8.68  &  1.39  &  -0.41  &  4.6  &  -0.18  &  53.62  &  -0.07  &  -1.42  &  -0.120 \pm 0.012  \\ 
9196-12701  &  8.61  &  10.65  &  0.81  &  9.57  &  2.07  &  -0.39  &  2.7  &  -0.03  &  109.92  &  -0.11  &  -2.04  &  +0.03 \pm 0.07  \\ 
9871-12701  &  8.37  &  9.25  &  0.58  &  8.75  &  1.25  &  -1.28  &  3.1  &  -0.10  &  61.46  &  0.09  &  -1.53  &  -0.146 \pm 0.009  \\ 
9493-12701  &  8.49  &  9.55  &  0.55  &  8.96  &  1.80  &  -1.05  &  5.1  &  -0.15  &  13.10  &  -0.13  &  -1.60  &  -0.020 \pm 0.010  \\ 
8325-12701  &  8.41  &  10.00  &  0.43  &  8.71  &  1.31  &  -0.25  &  2.9  &  -0.19  &  31.07  &  -0.12  &  -1.25  &  -0.106 \pm 0.014  \\ 
8933-12701  &  8.54  &  10.49  &  0.78  &  9.27  &  2.59  &  0.23  &  4.1  &  -0.26  &  50.78  &  -0.04  &  -1.27  &  +0.02 \pm 0.03  \\ 
8078-12701  &  8.58  &  10.95  &  0.86  &  9.30  &  2.56  &  0.45  &  4.3  &  -0.19  &  105.61  &  -0.07  &  -1.50  &  +0.025 \pm 0.006  \\ 
8147-12701  &  8.41  &  9.93  &  0.49  &  8.76  &  1.69  &  -0.64  &  5.1  &  -0.18  &  28.40  &  -0.05  &  -1.57  &  -0.045 \pm 0.009  \\ 
8454-12701  &  8.38  &  10.19  &  0.34  &  8.53  &  1.67  &  0.34  &  5.8  &  -0.32  &  19.98  &  -0.04  &  -0.85  &  -0.046 \pm 0.005  \\ 
8455-12701  &  8.39  &  10.35  &  0.40  &  8.58  &  2.27  &  0.57  &  5.3  &  -0.30  &  51.85  &  -0.07  &  -0.78  &  -0.066 \pm 0.004  \\ 
9185-12701  &  \mbox{---}   &  10.39  &  0.64  &  8.75  &  1.34  &  -0.02  &  5.5  &  -0.14  &  12.88  &  \mbox{---}   &  -1.42  &  +0.015 \pm 0.009  \\ 
8149-12701  &  8.38  &  10.15  &  0.31  &  8.48  &  1.58  &  0.20  &  6.3  &  -0.13  &  12.76  &  0.11  &  -0.95  &  -0.017 \pm 0.007  \\ 
8257-12701  &  8.52  &  10.63  &  0.56  &  8.89  &  2.47  &  0.91  &  5.4  &  -0.26  &  59.22  &  -0.01  &  -0.72  &  -0.023 \pm 0.001  \\ 
8611-12701  &  8.40  &  9.89  &  0.46  &  8.79  &  1.64  &  -0.31  &  4.4  &  -0.32  &  13.10  &  -0.10  &  -1.20  &  -0.081 \pm 0.005  \\ 
8984-12701  &  8.35  &  9.66  &  0.40  &  8.85  &  1.42  &  -0.89  &  5.9  &  -0.25  &  16.30  &  -0.04  &  -1.54  &  -0.139 \pm 0.012  \\ 
8259-12701  &  8.29  &  9.51  &  0.39  &  8.61  &  1.60  &  -0.87  &  5.9  &  -0.35  &  13.04  &  -0.04  &  -1.38  &  -0.117 \pm 0.007  \\ 
8655-12701  &  8.39  &  9.76  &  0.49  &  8.70  &  1.45  &  -0.70  &  4.9  &  -0.25  &  13.01  &  -0.10  &  -1.46  &  -0.059 \pm 0.005  \\ 
8613-12701  &  8.58  &  10.52  &  0.76  &  9.11  &  2.49  &  0.41  &  4.3  &  -0.23  &  262.46  &  -0.03  &  -1.11  &  -0.009 \pm 0.014  \\ 
8442-12701  &  8.45  &  10.53  &  0.54  &  8.72  &  2.24  &  0.93  &  4.5  &  -0.36  &  113.06  &  -0.01  &  -0.60  &  -0.029 \pm 0.007  \\ 
8252-12701  &  8.23  &  9.90  &  0.27  &  8.85  &  -0.71  &  -0.87  &  4.3  &  -0.23  &  166.22  &  -0.13  &  -1.76  &  -0.12 \pm 0.04  \\ 
9028-12701  &  8.59  &  10.49  &  0.66  &  9.22  &  2.11  &  -0.03  &  3.5  &  -0.12  &  56.25  &  -0.08  &  -1.52  &  +0.050 \pm 0.017  \\ 
7962-12701  &  8.40  &  9.96  &  0.47  &  8.81  &  1.51  &  -0.51  &  4.7  &  -0.46  &  16.36  &  -0.05  &  -1.48  &  -0.081 \pm 0.005  \\ 
7958-12701  &  8.29  &  9.61  &  0.20  &  8.46  &  1.60  &  -0.24  &  5.1  &  -0.25  &  17.54  &  -0.04  &  -0.85  &  -0.055 \pm 0.010  \\ 
8727-12701  &  8.44  &  10.18  &  0.56  &  8.99  &  1.74  &  -0.33  &  5.0  &  -0.08  &  13.03  &  -0.35  &  -1.52  &  +0.002 \pm 0.008  \\ 
8547-12701  &  8.50  &  11.07  &  0.81  &  9.57  &  2.65  &  1.17  &  3.5  &  -0.10  &  150.86  &  0.12  &  -0.89  &  +0.040 \pm 0.016  \\ 
8138-12701  &  8.44  &  10.23  &  0.44  &  8.56  &  1.50  &  0.10  &  5.3  &  -0.31  &  13.96  &  -0.08  &  -1.13  &  -0.031 \pm 0.008  \\ 
9501-12701  &  8.56  &  10.07  &  0.66  &  9.14  &  1.57  &  -0.75  &  3.6  &  -0.42  &  12.92  &  -0.17  &  -1.81  &  -0.01 \pm 0.06  \\ 
7815-12701  &  8.33  &  9.43  &  0.42  &  8.70  &  1.59  &  -1.06  &  4.0  &  -0.33  &  13.08  &  0.06  &  -1.49  &  -0.067 \pm 0.010  \\ 
8949-12701  &  \mbox{---}   &  9.54  &  0.64  &  9.47  &  0.81  &  -1.40  &  1.7  &  -0.06  &  12.96  &  \mbox{---}   &  -1.93  &  +0.03 \pm 0.03  \\ 
8326-12701  &  8.55  &  10.58  &  0.84  &  9.46  &  2.47  &  0.08  &  4.1  &  -0.10  &  121.11  &  -0.03  &  -1.50  &  +0.006 \pm 0.010  \\ 
9881-12701  &  8.52  &  10.55  &  0.49  &  8.66  &  2.01  &  0.78  &  5.1  &  -0.17  &  32.36  &  -0.13  &  -0.76  &  -0.023 \pm 0.003  \\ 
8456-12701  &  8.30  &  9.88  &  0.45  &  8.61  &  1.29  &  -0.53  &  5.2  &  -0.29  &  20.05  &  -0.24  &  -1.41  &  -0.070 \pm 0.008  \\ 
8131-12701  &  8.29  &  9.82  &  0.39  &  8.64  &  1.66  &  -0.52  &  5.9  &  -0.40  &  48.77  &  -0.06  &  -1.34  &  -0.099 \pm 0.003  \\ 
10001-12701  &  8.39  &  9.99  &  0.40  &  8.63  &  2.13  &  -0.13  &  5.1  &  -0.22  &  69.22  &  -0.02  &  -1.12  &  -0.062 \pm 0.013  \\ 
9031-12701  &  8.46  &  10.10  &  0.52  &  8.90  &  1.78  &  -0.25  &  4.0  &  -0.15  &  15.15  &  -0.14  &  -1.34  &  -0.064 \pm 0.019  \\ 
8452-12701  &  8.38  &  9.96  &  0.44  &  8.58  &  1.34  &  -0.40  &  4.9  &  -0.20  &  11.90  &  -0.08  &  -1.36  &  -0.060 \pm 0.017  \\ 
8084-12701  &  8.34  &  10.15  &  0.39  &  8.68  &  1.69  &  -0.24  &  5.7  &  -0.35  &  21.89  &  -0.02  &  -1.39  &  -0.121 \pm 0.007  \\ 
8085-12701  &  8.32  &  10.43  &  0.42  &  8.59  &  1.92  &  0.65  &  6.4  &  -0.32  &  19.22  &  -0.04  &  -0.78  &  -0.059 \pm 0.003  \\ 
8726-12701  &  8.55  &  10.54  &  0.66  &  9.27  &  1.96  &  0.06  &  5.4  &  -0.08  &  16.18  &  -0.08  &  -1.47  &  +0.038 \pm 0.008  \\ 
8146-12701  &  8.59  &  10.57  &  0.65  &  9.26  &  2.17  &  -0.03  &  3.6  &  -0.11  &  12.94  &  -0.07  &  -1.60  &  +0.024 \pm 0.016  \\ 
8453-12701  &  8.58  &  10.37  &  0.63  &  9.27  &  2.07  &  -0.26  &  3.4  &  -0.18  &  12.97  &  -0.04  &  -1.63  &  +0.025 \pm 0.018  \\ 
8600-12701  &  8.48  &  10.21  &  0.64  &  9.07  &  1.61  &  -0.49  &  5.2  &  -0.15  &  18.44  &  -0.06  &  -1.70  &  -0.013 \pm 0.009  \\ 
8445-12701  &  8.59  &  10.85  &  0.78  &  9.20  &  2.30  &  0.49  &  4.8  &  -0.22  &  69.26  &  -0.04  &  -1.36  &  +0.050 \pm 0.005  \\ 
8082-12701  &  8.53  &  10.48  &  0.64  &  9.08  &  2.06  &  0.13  &  3.5  &  -0.19  &  47.50  &  -0.10  &  -1.35  &  -0.010 \pm 0.005  \\ 
8465-12701  &  8.53  &  10.33  &  0.53  &  8.92  &  2.03  &  0.41  &  5.0  &  -0.36  &  26.56  &  -0.11  &  -0.92  &  +0.012 \pm 0.004  \\ 
8602-12701  &  \mbox{---}   &  10.65  &  0.84  &  9.18  &  2.56  &  -0.02  &  1.3  &  -0.08  &  75.04  &  \mbox{---}   &  -1.68  &  +0.06 \pm 0.04  \\ 
8141-12701  &  8.51  &  10.61  &  0.66  &  9.15  &  2.38  &  0.47  &  4.4  &  -0.17  &  68.47  &  -0.07  &  -1.14  &  -0.003 \pm 0.005  \\ 
8446-12701  &  8.25  &  9.49  &  0.34  &  8.50  &  1.29  &  -0.59  &  6.0  &  -0.16  &  39.89  &  -0.01  &  -1.08  &  -0.117 \pm 0.006  \\ 
9486-12701  &  \mbox{---}   &  10.96  &  0.75  &  \mbox{---}   &  \mbox{---}   &  0.29  &  4.6  &  \mbox{---}   &  89.71  &  \mbox{---}   &  -1.67  &  +0.00 \pm 0.03  \\ 
8604-12701  &  8.58  &  10.93  &  0.75  &  9.48  &  2.10  &  -0.03  &  4.4  &  -0.07  &  110.94  &  -0.03  &  -1.96  &  -0.01 \pm 0.03  \\ 
9035-12701  &  8.38  &  9.92  &  0.26  &  8.40  &  1.50  &  0.27  &  5.4  &  -0.23  &  15.35  &  -0.09  &  -0.65  &  -0.061 \pm 0.009  \\ 
8991-12701  &  8.41  &  9.92  &  0.56  &  9.04  &  1.83  &  -0.47  &  3.6  &  -0.06  &  16.04  &  -0.08  &  -1.39  &  -0.06 \pm 0.03  \\ 
8931-12701  &  8.45  &  10.30  &  0.45  &  8.87  &  1.65  &  0.13  &  4.3  &  -0.31  &  41.19  &  -0.10  &  -1.17  &  -0.026 \pm 0.010  \\ 
8983-12701  &  8.55  &  10.35  &  0.68  &  9.18  &  2.10  &  -0.02  &  3.7  &  -0.27  &  73.57  &  -0.03  &  -1.37  &  +0.019 \pm 0.009  \\ 
8462-12701  &  8.39  &  9.82  &  0.45  &  8.76  &  1.69  &  -0.49  &  5.5  &  -0.25  &  13.01  &  -0.04  &  -1.31  &  -0.061 \pm 0.005  \\ 
8724-12701  &  8.56  &  10.56  &  0.68  &  9.24  &  2.05  &  0.14  &  4.3  &  -0.18  &  34.13  &  -0.12  &  -1.42  &  +0.002 \pm 0.008  \\ 
8332-12701  &  8.42  &  10.19  &  0.38  &  8.70  &  1.83  &  0.18  &  4.9  &  -0.23  &  16.27  &  -0.09  &  -1.01  &  -0.017 \pm 0.005  \\ 
8464-12701  &  8.55  &  10.12  &  0.60  &  9.12  &  1.71  &  -0.29  &  4.9  &  -0.46  &  12.99  &  -0.06  &  -1.41  &  +0.029 \pm 0.010  \\ 
8258-12701  &  8.44  &  10.05  &  0.52  &  8.84  &  1.68  &  -0.47  &  4.1  &  -0.24  &  14.45  &  -0.11  &  -1.52  &  -0.051 \pm 0.008  \\ 
8320-12701  &  8.41  &  9.92  &  0.45  &  8.72  &  1.82  &  -0.28  &  3.4  &  -0.15  &  22.67  &  -0.09  &  -1.20  &  -0.085 \pm 0.019  \\ 
9034-12701  &  8.29  &  9.24  &  0.44  &  8.58  &  1.14  &  -0.93  &  7.1  &  -0.23  &  67.37  &  -0.11  &  -1.18  &  -0.105 \pm 0.005  \\ 
\enddata
\tablecomments{From left to right the columns correspond to: (1) galaxy ID defined as the [plate]-[ifudesign] of the MaNGA observations; (2) gas-phase oxygen abundance value at $R_{e}$; (3) galaxy mass; (4) $g-r$ color; (5) average LW stellar age at $R_{e}$; (6) average stellar mass density at $R_{e}$; (7) integrated SFR; (8) morphological type; (9) average LW stellar metallicity at $R_{e}$; (10) central velocity dispersion; (11) slope of the oxygen abundance gradient; (12) sSFR; and (13) slope of the local $\rm SFR-Z_g$ relation.}
\tablecomments{Table \ref{tab:app1} is published in its entirety in the machine readable format.  Only a portion containing 100 galaxies is shown here for guidance regarding its form and content.}
\end{deluxetable*}
\end{longrotatetable}

\bibliographystyle{aasjournal}
\bibliography{bibliography}

\end{document}